\documentclass[hyperref]{ctexart}
\usepackage[left=2.50cm, right=2.50cm, top=2.50cm, bottom=2.50cm]{geometry} 
\usepackage{helvet}
\usepackage{amsmath, amsfonts, amssymb} 
\usepackage[english]{babel}
\usepackage{graphicx}   
\usepackage{url}        
\usepackage{bm}         
\usepackage{multirow}
\usepackage{booktabs}
\usepackage{algorithm}
\usepackage{algorithmic}
\usepackage{float}
\usepackage{epstopdf}
\usepackage{hyperref}
\newtheorem{remark}{Remark}
\newcounter{RomanNumber}
\newcommand{\MyRoman}[1]{\setcounter{RomanNumber}{#1}\Roman{RomanNumber}}

\usepackage{fancyhdr} 
\usepackage{hyperref} 
\hypersetup{colorlinks, bookmarks, unicode} 

\title{\textbf{EM-Type Algorithms for DOA Estimation in Unknown Nonuniform Noise}}

\author{\sffamily{Ming-yan Gong}\footnote{M. Gong is with the School of Information and Electronics, Beijing Institute of Technology, Beijing 100081, China (e-mail: \href{mailto:jinyan0_o@outlook.com}{jinyan0\_o@outlook.com}).}\text{~~and~}\sffamily{Bin Lyu}\footnote{B. Lyu is with the Key Laboratory of Ministry of Education in Broadband Wireless Communication and Sensor Network Technology, Nanjing University of Posts and Telecommunications, Nanjing 210003, China (e-mail: \href{mailto:blyu@njupt.edu.cn}{blyu@njupt.edu.cn}).}}

\date{}

\begin{document}

\maketitle

\noindent{\bf Abstract: }The expectation--maximization (EM) algorithm updates all of the parameter estimates simultaneously, which is not applicable to direction of arrival (DOA) estimation in unknown nonuniform noise. In this work, we present several efficient EM-type algorithms, which update the parameter estimates sequentially, for solving both the deterministic and stochastic maximum--likelihood (ML) direction finding problems in unknown nonuniform noise. Specifically, we design a generalized EM (GEM) algorithm and a space-alternating generalized EM (SAGE) algorithm for computing the deterministic ML estimator. Simulation results show that the SAGE algorithm outperforms the GEM algorithm. Moreover, we design two SAGE algorithms for computing the stochastic ML estimator, in which the first updates the DOA estimates simultaneously while the second updates the DOA estimates sequentially. Simulation results show that the second SAGE algorithm outperforms the first one.\\
		


\section{Introduction}

Estimating the directions of arrival (DOAs) of narrowband far-field sources using sensor arrays is an important task in array signal processing and several types of estimation methods have been proposed in the literature \cite{bibitem1}, \cite{bibitem2}. In particular, the classic maximum--likelihood (ML) method plays an essential role \cite{bibitem3}--\cite{bibitem5}. However, ML direction finding problems are non-convex and we are hard to derive their solutions in closed form.

For obtaining ML estimates in DOA estimation efficiently, the expectation--maximization (EM) algorithm in \cite{bibitem6} has been adopted to solve ML direction finding problems \cite{bibitem7}, \cite{bibitem8}. Further, the space-alternating generalized EM (SAGE) algorithm proposed in \cite{bibitem9} has been also applied to DOA estimation in order to speed up the convergence of the EM algorithm \cite{bibitem10}--\cite{bibitem12}. However, these EM and SAGE algorithms are usually derived under the known or unknown uniform noise model, i.e., the sensor noise variances must be equal, which may be impractical in certain applications. Hence, the two algorithms should be developed for DOA estimation in the presence of other noise models.

As a general case of the uniform noise model, nonuniform noise has attracted increasing attention. Nonuniform noise is with an arbitrary diagonal covariance matrix, i.e., the sensor noise variances can be unequal. Obviously, classical eigenstructure based subspace methods, e.g., the MUSIC algorithm, cannot be \emph{directly} employed for DOA estimation in unknown nonuniform noise. For applying subspace methods to DOA estimation in nonuniform noise, a subspace separation approach is proposed in \cite{bibitem13} while in \cite{bibitem14}, the noise covariance matrix is first estimated and the sensor data are then prewhitened. Moreover, the signal subspace can be determined by maximizing the stochastic log-likelihood function (LLF) or solving a least-squares problem \cite{bibitem15}. In \cite{bibitem16}, the signal and noise subspaces are separated by means of the eigendecomposition of a reduced array covariance matrix when the sources are uncorrelated. The authors in \cite{bibitem17} utilize the reduced array covariance matrix in \cite{bibitem16} and propose a non-iterative two-phase subspace-based DOA estimation method. By analyzing the array covariance matrix in nonuniform noise, an optimization problem based on the signal subspace is formulated in \cite{bibitem18}, which leads to a new DOA estimator.

The ML method can enjoy excellent statistical properties \cite{bibitem4}, \cite{bibitem5}. But, the analyses in \cite{bibitem19} and \cite{bibitem20} suggest that both the deterministic and stochastic ML direction finding problems in unknown nonuniform noise cannot be reduced to two problems with respect to only the DOAs because of the noise parameters and generally involve high-dimensional search algorithms. For efficiently computing both the deterministic and stochastic ML estimators, two alternating maximization (AM) algorithms are presented in \cite{bibitem19} and \cite{bibitem20}, respectively. Unfortunately, the two AM algorithms include high-dimensional numerical search at every iteration and are thus computationally intensive, which motivates this work.

In this work, we first try to apply the EM algorithm, each iteration of which consists of an expectation step (E-step) and a maximization step (M-step). At the M-step, however, the EM algorithm updates all of the parameter estimates simultaneously, which requires high-dimensional numerical search due to the unknown noise parameters. Hence, we present several efficient EM-type algorithms, which update the parameter estimates sequentially and only require low-dimensional numerical search at every iteration, for solving both the deterministic and stochastic ML direction finding problems in unknown nonuniform noise. Specifically, we design a generalized EM (GEM) algorithm based on the expectation-conditional maximization (ECM) algorithm \cite{bibitem21} and an SAGE algorithm, which updates the DOA estimates sequentially, for computing the deterministic ML estimator. Simulation results show that the SAGE algorithm outperforms the GEM algorithm, i.e., the SAGE algorithm converges faster and can avoid the convergence to an unwanted stationary point of the LLF more efficiently. Moreover, we design two SAGE algorithms for computing the stochastic ML estimator, in which the first updates the DOA estimates simultaneously while the second updates the DOA estimates sequentially. Simulation results show that the second SAGE algorithm outperforms the first one.

\emph{Notations:}
$(\cdot)^T$ and $(\cdot)^H$ denote transpose and conjugate transpose, respectively. $\boldsymbol{0}=[0~\cdots~0]^T$ and $\boldsymbol{1}=[1~\cdots~1]^T$. $\Vert\mathbf{v}\Vert$ and $[\mathbf{v}]_i$
denote the Euclidean norm and $i$th element of a vector $\mathbf{v}$, respectively. $\mathbf{M}^{-1}$, $\mathrm{Det}(\mathbf{M})$, $\mathrm{Tr}(\mathbf{M})$, and $[\mathbf{M}]_{i,i}$ denote the inversion, determinant, trace, $i$th diagonal element of a square matrix $\mathbf{M}$, respectively. $\mathbf{I}_N$ and $\mathbf{0}_N$ are the $N\times N$ identity and zero matrices, respectively. $\mathbf{M}\ge\mathbf{0}_N$ and $\mathbf{M}>\mathbf{0}_N$ denote that the $N\times N$ square matrix $\mathbf{M}$ is positive semi-definite and definite, respectively. $\mathbb{E}\{\cdot\}$ and $\mathbb{D}\{\cdot\}$ denote expectation and covariance, respectively. $\jmath$ is the imaginary unit.

\section{Data Model and Problem Formulation}

Let us consider an array of $N$ sensors, which receives the signals transmitted by $M$ narrowband far-field sources with the same known center wavelength $\lambda$. For simplicity, the array is assumed to be a uniform linear array of inter-sensor spacing $\lambda/2$. The DOA of the $m$th source is denoted by $\theta_m\in(0,\pi)$. Then, the signal vector received at this array can be written as \cite{bibitem1}, \cite{bibitem2}
\begin{equation}\label{eq1}     
\mathbf{v}(t)={\sum}_{m=1}^M\mathbf{d}(\theta_m)f_m(t)+\mathbf{z}(t)=\mathbf{D}(\boldsymbol{\theta})\mathbf{f}(t)+\mathbf{z}(t),
\end{equation}
where
\begin{enumerate}
\item $\mathbf{d}(\theta_m)=\big[1~e^{-\jmath\pi\cos(\theta_m)}~\cdots~e^{-\jmath(N-1)\pi\cos(\theta_m)}\big]^T$ is the steering vector of the $m$th source, 
\item $f_m(t)$ is the $m$th source signal received at the $1$st sensor and with power $P_m$,
\item $\mathbf{z}(t)\sim\mathcal{CN}\big(\mathbf{0},\boldsymbol{\Sigma}\big)$ is a white complex Gaussian noise vector with covariance matrix $\boldsymbol{\Sigma}=\mathrm{diag}\{\boldsymbol{\sigma}\}$ and $\boldsymbol{\sigma}=[\sigma_{1}~\cdots~\sigma_{N}]^T>\boldsymbol{0}$, i.e., the nonuniform noise model.
\end{enumerate}
When $\sigma_1=\cdots=\sigma_N=\sigma$, $\boldsymbol{\Sigma}=\sigma\mathbf{I}_N$ and the nonuniform noise model reduces to the uniform noise model. In \eqref{eq1}, $\boldsymbol{\theta}=[\theta_1~\cdots~\theta_M]^T\in\boldsymbol{\Psi}$ with $\boldsymbol{\Psi}=(0,\pi)^M$, $\mathbf{D}(\boldsymbol{\theta})=[\mathbf{d}(\theta_1)~\cdots~\mathbf{d}(\theta_M)]$, and $\mathbf{f}(t)=[f_1(t)~\cdots~f_M(t)]^T$.

In the EM algorithm, the complete data need to be defined. We thus sample the received signal and design the samples (``snapshots'') or incomplete data by \cite{bibitem7}, \cite{bibitem11}
\begin{eqnarray}\label{eq2}     
\mathbf{v}(t)={\sum}_{m=1}^M\big[\mathbf{d}(\theta_m)f_m(t)+\mathbf{z}_m(t)\big]
={\sum}_{m=1}^M\mathbf{g}_m(t), t=1,2,\dots,T,
\end{eqnarray}
where
\begin{enumerate}
\item $T$ is the total number of samples,
\item the $\mathbf{g}_m(t)$'s are the underlying complete data,
\item the $\mathbf{z}_m(t)$'s are mutually independent noise vectors and $\mathbf{z}_m(t)\sim\mathcal{CN}\big(\boldsymbol{0},\boldsymbol{\Sigma}_m\big)$ with $\boldsymbol{\Sigma}_m=\mathrm{diag}\{\boldsymbol{\sigma}_m\}$, $\boldsymbol{\sigma}_m=[\sigma_{1,m}~\cdots~\sigma_{N,m}]^T>\boldsymbol{0}$, and $\boldsymbol{\Sigma}=\sum_{m=1}^M\boldsymbol{\Sigma}_m$.
\end{enumerate}
When the noise is uniform, we set $\sigma_{1,m}=\cdots=\sigma_{N,m}=\delta_m$, i.e., $\boldsymbol{\Sigma}_m=\delta_m\mathbf{I}_N$ and $\sigma=\sum_{m=1}^M\delta_m$ \cite{bibitem22}.
Since both the incomplete- and complete-data LLFs must determine the statistical model of the $f_m(t)$'s, we consider the so-called deterministic and stochastic signal models separately.

\subsection{Deterministic Signal Model}

In the deterministic signal model \cite{bibitem7}, \cite{bibitem8}, \cite{bibitem19}, the $f_m(t)$'s are deterministic and unknown, so we have
$\mathbf{g}_m(t)\sim\mathcal{CN}\big(\mathbf{d}(\theta_m)f_m(t),\boldsymbol{\Sigma}_m\big)$ and 
$\mathbf{v}(t)\sim\mathcal{CN}\big(\mathbf{D}(\boldsymbol{\theta})\mathbf{f}(t),\boldsymbol{\Sigma}\big)$. Then, both the incomplete- and complete-data LLFs are, respectively, expressed by
\begin{subequations}
\begin{eqnarray}\label{eq3a}       
\mathcal{L}(\boldsymbol{\Phi},\boldsymbol{\sigma})=&\sum_{t=1}^T\ln\mathnormal{p}(\mathbf{v}(t);\boldsymbol{\theta},\mathbf{f}(t),\boldsymbol{\sigma})=-TN\ln(\pi)-T\sum_{n=1}^N\ln(\sigma_n)-\nonumber\\
&\sum_{t=1}^T\big[\mathbf{v}(t)-\mathbf{D}(\boldsymbol{\theta})\mathbf{f}(t)\big]^H\boldsymbol{\Sigma}^{-1}
\big[\mathbf{v}(t)-\mathbf{D}(\boldsymbol{\theta})\mathbf{f}(t)\big],
\end{eqnarray}
\begin{eqnarray}\label{eq3b}       
\mathcal{U}(\boldsymbol{\Phi},\boldsymbol{\Omega})
=&\sum_{t=1}^T\sum_{m=1}^M\ln\mathnormal{p}(\mathbf{g}_m(t);\theta_m,f_m(t),\boldsymbol{\sigma}_m)
=-TMN\ln(\pi)-T\sum_{m=1}^M\sum_{n=1}^N\ln(\sigma_{n,m})-\nonumber\\
&\sum_{t=1}^T\sum_{m=1}^M\big[\mathbf{g}_m(t)-\mathbf{d}(\theta_m)f_m(t)\big]^H
\boldsymbol{\Sigma}^{-1}_m\big[\mathbf{g}_m(t)-\mathbf{d}(\theta_m)f_m(t)\big],
\end{eqnarray}
\end{subequations}
where $\mathbf{F}=[\mathbf{f}(1)~\cdots~\mathbf{f}(T)]$, $\boldsymbol{\Phi}=(\boldsymbol{\theta},\mathbf{F})$ and $\boldsymbol{\Omega}=(\boldsymbol{\sigma}_1,\dots,\boldsymbol{\sigma}_M)$ denote the signal and noise parameters, respectively.
Based on \eqref{eq3a}, the deterministic ML direction finding problem is constructed as
\begin{eqnarray}\label{eq4}       
\max_{\boldsymbol{\theta}\in\boldsymbol{\Psi},\mathbf{F},\boldsymbol{\sigma}>\mathbf{0}}\mathcal{L}(\boldsymbol{\Phi},\boldsymbol{\sigma}).
\end{eqnarray}

\subsection{Stochastic Signal Model}

In the stochastic signal model, the source signals are modelled as $f_m(t)\sim\mathcal{CN}(0,P_m)$ where $P_m$ is the power of the $m$th source.
For simplicity, we assume that the sources are independent of the noise and all of the $f_m(t)$'s are mutually independent \cite{bibitem7}, \cite{bibitem8}. Moreover, let $\boldsymbol{\Sigma}_m=\alpha_m\boldsymbol{\Sigma}$ with $\boldsymbol{\alpha}=[\alpha_1~\cdots~\alpha_M]^T>\mathbf{0}$ and $\mathbf{1}^T\boldsymbol{\alpha}=1$ known. Then, we have
$\mathbf{g}_m(t)\sim\mathcal{CN}(\mathbf{0},\mathbf{H}_m)$ and 
$\mathbf{v}(t)\sim\mathcal{CN}(\mathbf{0},\mathbf{H}_v)$
where $\mathbf{H}_m=P_m\mathbf{d}(\theta_m)\mathbf{d}^H(\theta_m)+\alpha_m\boldsymbol{\Sigma}>\mathbf{0}_N$ and $\mathbf{H}_v=\sum_{m=1}^M\mathbf{H}_m=\sum_{m=1}^MP_m\mathbf{d}(\theta_m)\mathbf{d}^H(\theta_m)+\boldsymbol{\Sigma}>\mathbf{0}_N$. Both the incomplete- and complete-data LLFs are, respectively, expressed by
\begin{subequations}
\begin{eqnarray}\label{eq5a}       
\mathcal{L}(\boldsymbol{\Phi},\boldsymbol{\sigma})=\sum_{t=1}^T\ln\mathnormal{p}(\mathbf{v}(t);\boldsymbol{\theta},\mathbf{P},\boldsymbol{\sigma})
=-T\big[N\ln(\pi)+\ln\big(\mathrm{Det}(\mathbf{H}_v)\big)+\mathrm{Tr}\big(\mathbf{H}^{-1}_v\hat{\mathbf{R}}_v\big)\big],
\end{eqnarray}
\begin{eqnarray}\label{eq5b}       
\mathcal{U}(\boldsymbol{\Phi},\boldsymbol{\sigma})
=\sum_{t=1}^T\sum_{m=1}^M\ln\mathnormal{p}(\mathbf{g}_m(t);\theta_m,P_m,\boldsymbol{\sigma})
=-T\sum_{m=1}^M\big[N\ln(\pi)+\ln\big(\mathrm{Det}(\mathbf{H}_m)\big)+\mathrm{Tr}\big(\mathbf{H}^{-1}_m\hat{\mathbf{R}}_m\big)\big],
\end{eqnarray}
\end{subequations}
where $\mathbf{P}=[P_1~\cdots~P_M]^T$, $\boldsymbol{\Phi}=(\boldsymbol{\theta},\mathbf{P})$, $\hat{\mathbf{R}}_v=(1/T)\sum_{t=1}^T\mathbf{v}(t)\mathbf{v}^H(t)$, and $\hat{\mathbf{R}}_m=(1/T)\sum_{t=1}^T\mathbf{g}_m(t)\mathbf{g}^H_m(t)$.
Based on \eqref{eq5a}, the stochastic ML direction finding problem is constructed as 
\begin{eqnarray}\label{eq6}       
\max_{\boldsymbol{\theta}\in\boldsymbol{\Psi},\mathbf{P}\ge\mathbf{0},\boldsymbol{\sigma}>\mathbf{0}}\mathcal{L}(\boldsymbol{\Phi},\boldsymbol{\sigma}).
\end{eqnarray}

\begin{remark}
Although problems \eqref{eq4} and \eqref{eq6} can be further reduced to some problems with respect to fewer unknown parameters \cite{bibitem19}, \cite{bibitem20}, these problems are still non-convex and high-dimensional, i.e., applying conventional gradient-type search algorithms to solve these problems is computationally expensive. For efficiently computing both the deterministic and stochastic ML estimators of $\boldsymbol{\theta}$ in \eqref{eq4} and \eqref{eq6}, we design some appropriate EM-type algorithms in the next two sections.
\end{remark}

\section{Deterministic Signal Model}

In this section, we present a GEM algorithm and an SAGE algorithm for solving problem \eqref{eq4}. For convenience, let $(\cdot)^{(b)}$ denote an iterative value at the $b$th iteration and $(\cdot)^{(0)}$ is an initial estimate.

\subsection{GEM Algorithm}

We first try to apply the EM algorithm and the E- and M-steps at the $b$th iteration are introduced below

\subsubsection{E-step}

The EM algorithm calculates the conditional expectation of the complete-data LLF in \eqref{eq3b} \cite{bibitem7}, i.e.,
\begin{eqnarray}\label{eq7}       
\mathbb{E}\left\{\mathcal{U}(\boldsymbol{\Phi},\boldsymbol{\Omega})\big|\mathbf{V};\boldsymbol{\Phi}^{(b-1)},\boldsymbol{\Omega}^{(b-1)}\right\}
=-TMN\ln(\pi)-T\sum_{m=1}^M\sum_{n=1}^N\ln(\sigma_{n,m})-\nonumber\\
\sum_{t=1}^T\sum_{m=1}^M\Big(\mathrm{Tr}\big(\boldsymbol{\Sigma}_m^{-1}\mathbf{G}^{(b)}_m\big)+\big[\mathbf{g}^{(b)}_m(t)-
\mathbf{d}(\theta_m)f_m(t)\big]^H\boldsymbol{\Sigma}^{-1}_m\big[\mathbf{g}^{(b)}_m(t)-\mathbf{d}(\theta_m)f_m(t)\big]\Big),
\end{eqnarray}
where $\mathbf{V}=[\mathbf{v}(1)~\cdots~\mathbf{v}(T)]$, the conditional probability density function of $\mathbf{g}_m(t)$ can be derived from \cite{bibitem23}, and
\begin{subequations}
\begin{eqnarray}       
\mathbf{g}^{(b)}_m(t)=&\mathbb{E}\left\{\mathbf{g}_m(t)\big|\mathbf{V};\boldsymbol{\Phi}^{(b-1)},\boldsymbol{\Omega}^{(b-1)}\right\}\nonumber\\
=&\mathbf{d}\big(\theta^{(b-1)}_m\big)f^{(b-1)}_m(t)+\boldsymbol{\Sigma}^{(b-1)}_m\big[\boldsymbol{\Sigma}^{(b-1)}\big]^{-1}
\big[\mathbf{v}(t)-\mathbf{D}(\boldsymbol{\theta}^{(b-1)})\mathbf{f}^{(b-1)}(t)\big],\forall m,t,
\end{eqnarray}
\begin{eqnarray}       
\mathbf{G}^{(b)}_m=\mathbb{D}\left\{\mathbf{g}_m(t)\big|\mathbf{V};\boldsymbol{\Phi}^{(b-1)},\boldsymbol{\Omega}^{(b-1)}\right\}
=\boldsymbol{\Sigma}^{(b-1)}_m-\boldsymbol{\Sigma}^{(b-1)}_m\big[\boldsymbol{\Sigma}^{(b-1)}\big]^{-1}\boldsymbol{\Sigma}^{(b-1)}_m,\forall m,t.
\end{eqnarray}
\end{subequations}

\subsubsection{M-step}

The EM algorithm estimates $\boldsymbol{\Phi}=(\boldsymbol{\theta},\mathbf{F})$ and $\boldsymbol{\Omega}=(\boldsymbol{\sigma}_1,\dots,\boldsymbol{\sigma}_M)$ by maximizing \eqref{eq7}, which results in the $M$ parallel subproblems:
\begin{eqnarray}\label{eq9}       
\min_{\theta_m\in(0,\pi),\mathbf{f}_m,\boldsymbol{\sigma}_m>\mathbf{0}}\sum_{n=1}^N\big[\ln(\sigma_{n,m})+\frac{c^{(b)}_{n,m}}{\sigma_{n,m}}\big]+\frac{1}{T}\sum_{t=1}^T\big[\mathbf{g}^{(b)}_m(t)-
\mathbf{d}(\theta_m)f_m(t)\big]^H\boldsymbol{\Sigma}^{-1}_m\big[\mathbf{g}^{(b)}_m(t)-\mathbf{d}(\theta_m)f_m(t)\big],\forall m,
\end{eqnarray}
where $\mathbf{f}_m=[f_m(1)~\cdots~f_m(T)]$ and $c^{(b)}_{n,m}=\big[\mathbf{G}^{(b)}_m\big]_{n,n}=\sigma^{(b-1)}_{n,m}\big(1-\sigma^{(b-1)}_{n,m}/\sigma^{(b-1)}_n\big)$ with $\sigma^{(b-1)}_n=\sum_{m=1}^M\sigma^{(b-1)}_{n,m}$.

Subproblems \eqref{eq9} are hard to be reduced to parallel subproblems each with respect to only one exclusive DOA, we thus resort to the ECM algorithm \cite{bibitem21} and replace the M-step with the following two conditional maximization steps (CM-steps), i.e., the EM algorithm becomes the ECM algorithm.

\subsubsection{CM-steps}

At the first CM-step, the ECM algorithm estimates $\boldsymbol{\Phi}$ but holds $\boldsymbol{\Omega}=\boldsymbol{\Omega}^{(b-1)}$ fixed. Then, subproblems \eqref{eq9} are reduced to
\begin{eqnarray}\label{eq10}       
\min_{\theta_m\in(0,\pi),\mathbf{f}_m}\frac{1}{T}\sum_{t=1}^T\big\Vert\tilde{\mathbf{g}}^{(b)}_m(t)-\tilde{\mathbf{d}}(\theta_m)f_m(t)\big\Vert^2,\forall m,
\end{eqnarray}
where $\boldsymbol{\Sigma}_m^{-1/2}=\mathrm{diag}\big\{1/\sqrt{\sigma_{1,m}},\dots,1/\sqrt{\sigma_{N,m}}\big\}$,
$\tilde{\mathbf{g}}^{(b)}_m(t)=\big[\boldsymbol{\Sigma}^{(b-1)}_m\big]^{-1/2}\mathbf{g}^{(b)}_m(t)$, and
$\tilde{\mathbf{d}}(\theta_m)=\big[\boldsymbol{\Sigma}^{(b-1)}_m\big]^{-1/2}\mathbf{d}(\theta_m)$.
Subproblems \eqref{eq10} can be solved in a separable manner and reduced to $M$ parallel one-dimensional search subproblems each with respect to only one exclusive DOA \cite{bibitem8}. Accordingly, the estimate of $\boldsymbol{\Phi}$ is updated by
\begin{subequations}
\begin{eqnarray}\label{eq11a}       
\theta^{(b)}_m=\arg\max_{\theta_m\in(0,\pi)}\tilde{\mathbf{d}}^H(\theta_m)\widetilde{\mathbf{R}}^{(b)}_m\tilde{\mathbf{d}}(\theta_m),\forall m,
\end{eqnarray}
\begin{eqnarray}\label{eq11b}       
f^{(b)}_m(t)=\tilde{\mathbf{d}}^H\big(\theta^{(b)}_m\big)\tilde{\mathbf{g}}^{(b)}_m(t)/q^{(b)}_m,\forall m,t,
\end{eqnarray}
\end{subequations}
where $\widetilde{\mathbf{R}}^{(b)}_m=(1/T)\sum_{t=1}^T\tilde{\mathbf{g}}^{(b)}_m(t)\big[\tilde{\mathbf{g}}^{(b)}_m(t)\big]^H$ and
$q^{(b)}_m=\tilde{\mathbf{d}}^H\big(\theta^{(b)}_m\big)\tilde{\mathbf{d}}\big(\theta^{(b)}_m\big)=\sum_{n=1}^N\big[\sigma^{(b-1)}_{n,m}\big]^{-1}$.

At the second CM-step, the ECM algorithm estimates $\boldsymbol{\Omega}$ but holds $\boldsymbol{\Phi}=\boldsymbol{\Phi}^{(b)}$ fixed. Then, subproblems \eqref{eq9} are reduced to
\begin{eqnarray}\label{eq12}       
\min_{\boldsymbol{\sigma}_m>\mathbf{0}}\sum_{n=1}^N\big[\ln(\sigma_{n,m})+\frac{c^{(b)}_{n,m}+d^{(b)}_{n,m}}{\sigma_{n,m}}\big],\forall m,
\end{eqnarray}
where $\vert\cdot\vert$ denotes the modulus of a complex number and
\begin{eqnarray}       
d^{(b)}_{n,m}=\frac{1}{T}\sum_{t=1}^T\Big\vert\big[\mathbf{g}^{(b)}_m(t)-
\mathbf{d}\big(\theta^{(b)}_m\big)f^{(b)}_m(t)\big]_n\Big\vert^2\ge0,\forall n,m.\nonumber
\end{eqnarray}
Thus, the estimate of $\boldsymbol{\Omega}$ is updated by
\begin{eqnarray}\label{eq13}       
\sigma^{(b)}_{n,m}=c^{(b)}_{n,m}+d^{(b)}_{n,m},\forall n,m.
\end{eqnarray}

Unfortunately, simulation results show that the operation of the algorithm may not be smooth when the estimate of $\boldsymbol{\Omega}$ is updated by \eqref{eq13}. To ensure the stability, we decrease the difference between $\sigma^{(b)}_{n,m}$ and $\sigma^{(b-1)}_{n,m}$ by modifying \eqref{eq13} as
\begin{eqnarray}\label{eq14}          
\sigma^{(b)}_{n,m}=\beta\sigma^{(b-1)}_{n,m}+(1-\beta)\big(c^{(b)}_{n,m}+d^{(b)}_{n,m}\big),\forall n,m,
\end{eqnarray}
where $\beta\in[0,1]$ is adjusted in simulation.
Note that \eqref{eq14} guarantees the following monotonicity
\begin{eqnarray}
\ln\big(\sigma^{(b)}_{n,m}\big)+\frac{c^{(b)}_{n,m}+d^{(b)}_{n,m}}{\sigma^{(b)}_{n,m}}
\le~\ln\big(\sigma^{(b-1)}_{n,m}\big)+\frac{c^{(b)}_{n,m}+d^{(b)}_{n,m}}{\sigma^{(b-1)}_{n,m}},\forall n,m.\nonumber
\end{eqnarray}
Moreover, if $\boldsymbol{\sigma}^{(b-1)}_m>\mathbf{0},\forall m$, we can obtain
$c^{(b)}_{n,m}=\sigma^{(b-1)}_{n,m}\big(1-\sigma^{(b-1)}_{n,m}/\sigma^{(b-1)}_n\big)>0,\forall n,m,$ 
and
$\sigma^{(b)}_{n,m}\ge\beta\sigma^{(b-1)}_{n,m}+(1-\beta)c^{(b)}_{n,m}>0,\forall n,m,$ i.e., 
$\boldsymbol{\sigma}^{(b)}_m>\mathbf{0},\forall m.$

Although the M-step of the EM algorithm is replaced with the above two CM-steps of the ECM algorithm, we can easily prove that the monotonicity of GEM algorithms still holds, i.e.,
\begin{eqnarray}
\mathcal{L}\big(\boldsymbol{\Phi}^{(b)},\boldsymbol{\sigma}^{(b)}\big)\ge\mathcal{L}\big(\boldsymbol{\Phi}^{(b-1)},\boldsymbol{\sigma}^{(b-1)}\big).\nonumber
\end{eqnarray}
As stated in \cite{bibitem21}, the ECM algorithm is a GEM algorithm.

\subsection{SAGE Algorithm}\label{sub:32}

According to the above GEM algorithm, the corresponding SAGE algorithm is presented for speeding up the convergence. At every iteration, the SAGE algorithm sequentially updates the DOA estimates from $\theta_1$ to $\theta_M$ using $M$ cycles. For convenience, let $(\cdot)^{(b,i)}$ denote an iterative value at the $i$th cycle of the $b$th iteration, $(\cdot)^{(b,M)}=(\cdot)^{(b+1,0)}=(\cdot)^{(b)}$.

When the SAGE algorithm updates the estimate of $\theta_i$ at the $i$th cycle of the $b$th iteration, all of the noise is first allocated to the $i$th source signal component by \cite{bibitem9}--\cite{bibitem11}
\begin{equation}\label{eq15}          
\mathbf{g}_m(t)=
\left\{
\begin{array}{ll}
{\mathbf{d}(\theta_m)f_m(t)+\mathbf{z}(t)} & {m=i}, \\
{\mathbf{d}(\theta_m)f_m(t)} & {m\ne i}, \\
\end{array}
\right.
\end{equation}
which indicates that $\mathbf{g}_i(t)\sim\mathcal{CN}\big(\mathbf{d}(\theta_i)f_i(t),\boldsymbol{\Sigma}\big)$ and $\mathbf{g}_m(t)$ is deterministic for $m\ne i$. The corresponding complete-data LLF is written as
\begin{eqnarray}\label{eq16}       
\mathcal{U}(\theta_i,\mathbf{f}_i,\boldsymbol{\sigma})=\sum_{t=1}^T\ln\mathnormal{p}(\mathbf{g}_i(t);\theta_i,f_i(t),\boldsymbol{\sigma})
=-TN\ln(\pi)-T\sum_{n=1}^N\ln(\sigma_n)\nonumber\\-\sum_{t=1}^T\big[\mathbf{g}_i(t)-
\mathbf{d}(\theta_i)f_i(t)\big]^H\boldsymbol{\Sigma}^{-1}
\big[\mathbf{g}_i(t)-\mathbf{d}(\theta_i)f_i(t)\big].
\end{eqnarray}
The E- and CM-steps at the $i$th cycle of the $b$th iteration are introduced below.

\subsubsection{E-step}
Based on \eqref{eq16}, the SAGE algorithm calculates the conditional expectation of $\mathcal{U}(\theta_i,\mathbf{f}_i,\boldsymbol{\sigma})$, i.e.,
\begin{eqnarray}\label{eq17}       
\mathbb{E}\left\{\mathcal{U}(\theta_i,\mathbf{f}_i,\boldsymbol{\sigma})\big|\mathbf{V};\boldsymbol{\Phi}^{(b,i-1)},\boldsymbol{\sigma}^{(b,i-1)}\right\}
=-TN\ln(\pi)-T\sum_{n=1}^N\ln(\sigma_n)\nonumber\\-\sum_{t=1}^T\big[\mathbf{g}^{(b)}_i(t)-
\mathbf{d}(\theta_i)f_i(t)\big]^H
\boldsymbol{\Sigma}^{-1}\big[\mathbf{g}^{(b)}_i(t)-\mathbf{d}(\theta_i)f_i(t)\big],
\end{eqnarray}
where $\boldsymbol{\Phi}=(\boldsymbol{\theta},\mathbf{F})$ and 
\begin{subequations}
\begin{eqnarray}       
\mathbf{g}^{(b)}_i(t)=&\mathbf{g}^{(b,i)}_i(t)=\mathbb{E}\left\{\mathbf{g}_i(t)\big|\mathbf{V};\boldsymbol{\Phi}^{(b,i-1)},\boldsymbol{\sigma}^{(b,i-1)}\right\}\nonumber\\
=&\mathbf{d}\big(\theta^{(b,i-1)}_i\big)f^{(b,i-1)}_i(t)+
\big[\mathbf{v}(t)-\mathbf{D}\big(\boldsymbol{\theta}^{(b,i-1)}\big)\mathbf{f}^{(b,i-1)}(t)\big],\forall t,
\end{eqnarray}
\begin{eqnarray}       
\mathbb{D}\left\{\mathbf{g}_i(t)\big|\mathbf{V};\boldsymbol{\Phi}^{(b,i-1)},\boldsymbol{\sigma}^{(b,i-1)}\right\}=\mathbf{0}_N,\forall t.
\end{eqnarray}
\end{subequations}

\subsubsection{CM-steps}
The SAGE algorithm tries to estimate $\theta_i$, $\textbf{f}_i$, and $\boldsymbol{\sigma}$ by maximizing \eqref{eq17}, i.e.,
\begin{eqnarray}\label{eq19}       
\min_{\theta_i\in(0,\pi),\mathbf{f}_i,\boldsymbol{\sigma}>\mathbf{0}}\sum_{n=1}^N\ln(\sigma_n)+\frac{1}{T}\sum_{t=1}^T\big[\mathbf{g}^{(b)}_i(t)-
\mathbf{d}(\theta_i)f_i(t)\big]^H\boldsymbol{\Sigma}^{-1}\big[\mathbf{g}^{(b)}_i(t)-\mathbf{d}(\theta_i)f_i(t)\big].
\end{eqnarray}
However, problem \eqref{eq19} is hard to be reduced to a one-dimensional search problem with respect to only $\theta_i$, we use two CM-steps based on the ECM algorithm \cite{bibitem21}.

At the first CM-step, the SAGE algorithm estimates $\theta_i$ and $\mathbf{f}_i$ but holds $\boldsymbol{\sigma}=\boldsymbol{\sigma}^{(b,i-1)}$ fixed. Then, problem \eqref{eq19} is reduced to
\begin{eqnarray}\label{eq20}       
\min_{\theta_i\in(0,\pi),\mathbf{f}_i}\frac{1}{T}\sum_{t=1}^T\big\Vert\tilde{\mathbf{g}}^{(b)}_i(t)-\tilde{\mathbf{d}}(\theta_i)f_i(t)\big\Vert^2,
\end{eqnarray}
where 
$\tilde{\mathbf{g}}^{(b)}_i(t)=\big[\boldsymbol{\Sigma}^{(b,i-1)}\big]^{-1/2}\mathbf{g}^{(b)}_i(t)$ and 
$\tilde{\mathbf{d}}(\theta_i)=\big[\boldsymbol{\Sigma}^{(b,i-1)}\big]^{-1/2}\mathbf{d}(\theta_i)$.
Following \eqref{eq11a} and \eqref{eq11b}, the estimates of $\theta_i$ and $\mathbf{f}_i$ are updated by
\begin{subequations}
\begin{eqnarray}       
\theta^{(b)}_i=\theta^{(b,i)}_i=\arg\max_{\theta_i\in(0,\pi)}\tilde{\mathbf{d}}^H(\theta_i)\widetilde{\mathbf{R}}^{(b)}_i\tilde{\mathbf{d}}(\theta_i),
\end{eqnarray}
\begin{eqnarray}       
f^{(b)}_i(t)=f^{(b,i)}_i(t)=\tilde{\mathbf{d}}^H\big(\theta^{(b)}_i\big)\tilde{\mathbf{g}}^{(b)}_i(t)/q^{(b,i)},\forall t,
\end{eqnarray}
\end{subequations}
where $\widetilde{\mathbf{R}}^{(b)}_i=(1/T)\sum_{t=1}^T\tilde{\mathbf{g}}^{(b)}_i(t)\big[\tilde{\mathbf{g}}^{(b)}_i(t)\big]^H$ and 
$q^{(b,i)}=\tilde{\mathbf{d}}^H\big(\theta^{(b)}_i\big)\tilde{\mathbf{d}}\big(\theta^{(b)}_i\big)=\sum_{n=1}^N\big[\sigma^{(b,i-1)}_n\big]^{-1}$.

At the second CM-step, the SAGE algorithm estimates $\boldsymbol{\sigma}$ but holds $\theta_i=\theta^{(b)}_i$ and $\mathbf{f}_i=\mathbf{f}^{(b)}_i$ fixed. Then, problem \eqref{eq19} is reduced to
\begin{eqnarray}       
\min_{\boldsymbol{\sigma}>\mathbf{0}}\sum_{n=1}^N\big[\ln(\sigma_n)+\frac{d^{(b,i)}_n}{\sigma_n}\big],
\end{eqnarray}
where
$d^{(b,i)}_n=(1/T)\sum_{t=1}^T\big\vert\big[\mathbf{g}^{(b)}_i(t)-
\mathbf{d}\big(\theta^{(b)}_i\big)f^{(b)}_i(t)\big]_n\big\vert^2\ge0,\forall n.$
Thus, the estimate of $\boldsymbol{\sigma}$ is updated by
\begin{eqnarray}\label{eq23}       
\sigma^{(b,i)}_n=d^{(b,i)}_n,\forall n.
\end{eqnarray}

Unfortunately, the operation of the algorithm always is unstable in simulation when the estimate of $\boldsymbol{\sigma}$ is updated by \eqref{eq23}. Like \eqref{eq14}, we modify \eqref{eq23} as
\begin{eqnarray} \label{eq24}      
\sigma^{(b,i)}_n=\gamma\sigma^{(b,i-1)}_n+(1-\gamma)d^{(b,i)}_n,\forall n,
\end{eqnarray}
where $\gamma\in(0,1]$ is adjusted in simulation and tends to be larger than $\beta$. Note that in \eqref{eq24}, $\sigma^{(b,i)}_n\ge\gamma\sigma^{(b,i-1)}_n>0$ if $\sigma^{(b,i-1)}_n>0$, which ensures that if $\boldsymbol{\sigma}^{(b,i-1)}>\mathbf{0}$, we have
$\boldsymbol{\sigma}^{(b,i)}>\mathbf{0}.$

The other parameter estimates in $(\boldsymbol{\Phi}^{(b,i-1)},\boldsymbol{\sigma}^{(b,i-1)})$ are not updated at this cycle and their iterative values are denoted by
\begin{subequations}
\begin{eqnarray}       
\theta^{(b,i)}_m=\theta^{(b,i-1)}_m,\forall m\ne i,
\end{eqnarray}
\begin{eqnarray}       
f^{(b,i)}_m(t)=f^{(b,i-1)}_m(t),\forall m\ne i, t.
\end{eqnarray}
\end{subequations}
After this cycle, we have \cite{bibitem9}
\begin{eqnarray}
\mathcal{L}\big(\boldsymbol{\Phi}^{(b,i)},\boldsymbol{\sigma}^{(b,i)}\big)\ge\mathcal{L}\big(\boldsymbol{\Phi}^{(b,i-1)},\boldsymbol{\sigma}^{(b,i-1)}\big).
\end{eqnarray}
Thus, after the $b$th iteration, the monotonicity of the SAGE algorithm can be proved by
\begin{eqnarray}
\mathcal{L}\big(\boldsymbol{\Phi}^{(b)},\boldsymbol{\sigma}^{(b)}\big)=\mathcal{L}\big(\boldsymbol{\Phi}^{(b,M)},\boldsymbol{\sigma}^{(b,M)}\big)
\ge\cdots
\ge\mathcal{L}\big(\boldsymbol{\Phi}^{(b,0)},\boldsymbol{\sigma}^{(b,0)}\big)=\mathcal{L}\big(\boldsymbol{\Phi}^{(b-1)},\boldsymbol{\sigma}^{(b-1)}\big).
\end{eqnarray}

\section{Stochastic Signal Model}

In this section, we present two SAGE algorithms for solving problem \eqref{eq6}.
\subsection{First SAGE Algorithm}

The first SAGE algorithm simultaneously updates the DOA estimates based on the EM algorithm. To this end, we first try to use the EM algorithm and the E- and M-steps at the $b$th iteration are introduced below.

\subsubsection{E-step}

The EM algorithm calculates the conditional expectation of the complete-data LLF in \eqref{eq5b} \cite{bibitem7}, i.e.,
\begin{eqnarray}\label{eq28}       
\mathbb{E}\left\{\mathcal{U}(\boldsymbol{\Phi},\boldsymbol{\sigma})\big|\mathbf{V};\boldsymbol{\Phi}^{(b-1)},\boldsymbol{\sigma}^{(b-1)}\right\}
=-TMN\ln(\pi)-T\sum_{m=1}^M\big[\ln\big(\mathrm{Det}(\mathbf{H}_m)\big)+\mathrm{Tr}\big(\mathbf{H}^{-1}_m\hat{\mathbf{R}}^{(b)}_m\big)\big]
\end{eqnarray}
with $\boldsymbol{\Phi}=(\boldsymbol{\theta},\mathbf{P})$ and 
\begin{eqnarray}       
\hat{\mathbf{R}}^{(b)}_m=&\mathbb{E}\big\{\hat{\mathbf{R}}_m\big|\mathbf{V};\boldsymbol{\Phi}^{(b-1)},\boldsymbol{\sigma}^{(b-1)}\big\}\nonumber\\
=&\mathbf{H}^{(b-1)}_m\big[\mathbf{H}^{(b-1)}_v\big]^{-1}\hat{\mathbf{R}}_v\big[\mathbf{H}^{(b-1)}_v\big]^{-1}\mathbf{H}^{(b-1)}_m+
\mathbf{H}^{(b-1)}_m-\mathbf{H}^{(b-1)}_m\big[\mathbf{H}^{(b-1)}_v\big]^{-1}\mathbf{H}^{(b-1)}_m,\forall m,
\end{eqnarray}
where the conditional probability density function of $\mathbf{g}_m(t)$ can be derived from \cite{bibitem23} and
\begin{subequations}
\begin{eqnarray}       
\mathbb{E}\left\{\mathbf{g}_m(t)\big|\mathbf{V};\boldsymbol{\Phi}^{(b-1)},\boldsymbol{\sigma}^{(b-1)}\right\}
=\mathbf{H}^{(b-1)}_m\big[\mathbf{H}^{(b-1)}_v\big]^{-1}\mathbf{v}(t),\forall m,t,
\end{eqnarray}
\begin{eqnarray}       
\mathbb{D}\left\{\mathbf{g}_m(t)\big|\mathbf{V};\boldsymbol{\Phi}^{(b-1)},\boldsymbol{\sigma}^{(b-1)}\right\}
=\mathbf{H}^{(b-1)}_m-\mathbf{H}^{(b-1)}_m\big[\mathbf{H}^{(b-1)}_v\big]^{-1}\mathbf{H}^{(b-1)}_m\ge\mathbf{0}_N,\forall m,t.
\end{eqnarray}
\end{subequations}

\subsubsection{M-step}

The EM algorithm estimates $\boldsymbol{\Phi}$ and $\boldsymbol{\sigma}$ by maximizing \eqref{eq28}, i.e.,
\begin{eqnarray}\label{eq31}       
\begin{aligned}
\min_{\boldsymbol{\theta}\in\boldsymbol{\Psi},\mathbf{P}\ge\mathbf{0},\boldsymbol{\sigma}>\mathbf{0}}\sum_{m=1}^M\big[\ln\big(\mathrm{Det}(\mathbf{H}_m)\big)+\mathrm{Tr}\big(\mathbf{H}^{-1}_m\hat{\mathbf{R}}^{(b)}_m\big)\big],
\end{aligned}
\end{eqnarray}
which is hard to be reduced to parallel subproblems since all the $\mathbf{H}_m$'s are related to $\boldsymbol{\sigma}$.

To proceed, we hold $\boldsymbol{\sigma}=\boldsymbol{\sigma}^{(b-1)}$ fixed and reduce problem \eqref{eq31} to the $M$ parallel subproblems:
\begin{eqnarray}\label{eq32}       
\min_{\theta_m\in(0,\pi),P_m\ge0}\ln\big(\mathrm{Det}(\mathbf{K}_m)\big)+\mathrm{Tr}\big(\mathbf{K}^{-1}_m\widetilde{\mathbf{R}}^{(b)}_m\big),\forall m,
\end{eqnarray}
where
$\mathbf{H}_m=\big[\boldsymbol{\Sigma}^{(b-1)}\big]^{1/2}\mathbf{K}_m\big[\boldsymbol{\Sigma}^{(b-1)}\big]^{1/2}$,
$\mathbf{K}_m=P_m\tilde{\mathbf{d}}(\theta_m)\tilde{\mathbf{d}}^H(\theta_m)+\alpha_m\mathbf{I}_N$,
$\tilde{\mathbf{d}}(\theta_m)=\big[\boldsymbol{\Sigma}^{(b-1)}\big]^{-1/2}\mathbf{d}(\theta_m)$,
$\widetilde{\mathbf{R}}^{(b)}_m=\big[\boldsymbol{\Sigma}^{(b-1)}\big]^{-1/2}\hat{\mathbf{R}}^{(b)}_m\big[\boldsymbol{\Sigma}^{(b-1)}\big]^{-1/2}$.
Utilizing $\mathrm{Det}(\mathbf{K}_m)=\big(P_mq^{(b)}+\alpha_m\big)\alpha_m^{N-1}$ with $q^{(b)}=\Vert\tilde{\mathbf{d}}(\theta_m)\Vert^2=\sum_{n=1}^M\big[\sigma_n^{(b-1)}\big]^{-1}$ and
\begin{eqnarray}
\mathbf{K}_m^{-1}=\frac{1}{\alpha_m}\Big(\mathbf{I}_N-\frac{P_m\tilde{\mathbf{d}}(\theta_m)\tilde{\mathbf{d}}^H(\theta_m)}
{P_mq^{(b)}+\alpha_m}\Big),\nonumber
\end{eqnarray}
we further reduce subproblems \eqref{eq32} to
\begin{eqnarray}       
\min_{\theta_m\in(0,\pi),P_m\ge0}\ln\big(P_mq^{(b)}+\alpha_m\big)-
\frac{P_m\tilde{\mathbf{d}}^H(\theta_m)\widetilde{\mathbf{R}}^{(b)}_m\tilde{\mathbf{d}}(\theta_m)}{\alpha_m\big(P_mq^{(b)}+\alpha_m\big)},\forall m,
\end{eqnarray}
which can be solved in a separable manner and reduced to $M$ parallel one-dimensional search subproblems each with respect to only one exclusive DOA \cite{bibitem8}. Accordingly, the estimate of $\boldsymbol{\Phi}$ is updated by
\begin{subequations}
\begin{eqnarray}\label{eq34a}       
\theta^{(b)}_m=\arg\max_{\theta_m\in(0,\pi)}\tilde{\mathbf{d}}^H(\theta_m)\widetilde{\mathbf{R}}^{(b)}_m\tilde{\mathbf{d}}(\theta_m),\forall m,
\end{eqnarray}
\begin{eqnarray}\label{eq34b}       
P^{(b)}_m=\max\bigg\{\frac{1}{q^{(b)}}\Big(\frac{\tilde{\mathbf{d}}^H(\theta^{(b)}_m)\widetilde{\mathbf{R}}^{(b)}_m\tilde{\mathbf{d}}(\theta^{(b)}_m)}{q^{(b)}}
-\alpha_m\Big),0\bigg\},\forall m,
\end{eqnarray}
\end{subequations}
where $\theta^{(b)}_m$ is indeterminate if $P^{(b)}_m=0$. Then, we have
\begin{eqnarray}\label{eq35}     
\mathcal{L}\big(\boldsymbol{\Phi}^{(b)},\boldsymbol{\sigma}^{(b-1)}\big)\ge\mathcal{L}\big(\boldsymbol{\Phi}^{(b-1)},\boldsymbol{\sigma}^{(b-1)}\big).
\end{eqnarray}

\subsubsection{Additional E- and M-steps}\label{sec:413}

The above M-step does not update the estimate of $\boldsymbol{\sigma}$ due to the complexity in problem \eqref{eq31}, we thus add novel E- and M-steps to obtain $\boldsymbol{\sigma}^{(b)}$ easily, which leads to that the EM algorithm becomes the SAGE algorithm \cite{bibitem9}. To this end, we use $\mathbf{F}$ and $\mathbf{Z}=[\mathbf{z}(1)~\cdots~\mathbf{z}(T)]$ as the complete data. The corresponding complete-data LLF is written as
\begin{eqnarray}\label{eq36}       
\mathcal{U}(\mathbf{P},\boldsymbol{\sigma})=\sum_{t=1}^T\big[\ln\mathnormal{p}(\mathbf{f}(t);\mathbf{P})+\ln\mathnormal{p}(\mathbf{z}(t);\boldsymbol{\sigma})\big]
=-TM\ln(\pi)-T\sum_{m=1}^M\big[\ln(P_m)+\hat{P}_m/P_m\big]\nonumber \\
-TN\ln(\pi)-T\big[\ln\big(\mathrm{Det}(\boldsymbol{\Sigma})\big)+\mathrm{Tr}\big(\boldsymbol{\Sigma}^{-1}\hat{\mathbf{R}}_z\big)\big],
\end{eqnarray}
where $\hat{P}_m=(1/T)\sum_{t=1}^T\vert f_m(t)\vert^2$ and $\hat{\mathbf{R}}_z=(1/T)\sum_{t=1}^T\mathbf{z}(t)\mathbf{z}^H(t)$ are two complete-data sufficient statistics for $P_m$ and $\boldsymbol{\Sigma}$, respectively. Based on \eqref{eq36}, the additional E- and M-steps are introduced below.

At the additional E-step, the algorithm calculates the conditional expectation of $\mathcal{U}(\mathbf{P},\boldsymbol{\sigma})$, i.e.,
\begin{eqnarray}\label{eq37}       
\mathbb{E}\big\{\mathcal{U}(\mathbf{P},\boldsymbol{\sigma})\big|\mathbf{V};\boldsymbol{\Phi}^{(b)},\boldsymbol{\sigma}^{(b-1)}\big\}
=-TM\ln(\pi)-T\sum_{m=1}^M\big[\ln(P_m)+\hat{P}^{(b)}_m/P_m\big]\nonumber\\
-TN\ln(\pi)-T\big[\ln\big(\mathrm{Det}(\boldsymbol{\Sigma})\big)+\mathrm{Tr}\big(\boldsymbol{\Sigma}^{-1}\hat{\mathbf{R}}^{(b)}_z\big)\big]
\end{eqnarray}
with
\begin{subequations}
\begin{eqnarray}       
\hat{P}^{(b)}_m=\mathbb{E}\big\{\hat{P}_m\big|\mathbf{V};\boldsymbol{\Phi}^{(b)},\boldsymbol{\sigma}^{(b-1)}\big\}
=\big[\bar{\mathbf{q}}^{(b)}_m\big]^H\hat{\mathbf{R}}_v\bar{\mathbf{q}}^{(b)}_m+
P^{(b)}_m\big[1-\mathbf{d}^H\big(\theta_m^{(b)}\big)\bar{\mathbf{q}}^{(b)}_m\big]\ge0,\forall m,
\end{eqnarray}
\begin{eqnarray}\label{eq38b}       
\hat{\mathbf{R}}^{(b)}_z=&\mathbb{E}\big\{\hat{\mathbf{R}}_z\big|\mathbf{V};\boldsymbol{\Phi}^{(b)},\boldsymbol{\sigma}^{(b-1)}\big\}\nonumber\\
=&\boldsymbol{\Sigma}^{(b-1)}\big[\bar{\mathbf{H}}^{(b)}_v\big]^{-1}\hat{\mathbf{R}}_v\big[\bar{\mathbf{H}}^{(b)}_v\big]^{-1}\boldsymbol{\Sigma}^{(b-1)}+
\boldsymbol{\Sigma}^{(b-1)}-\boldsymbol{\Sigma}^{(b-1)}\big[\bar{\mathbf{H}}^{(b)}_v\big]^{-1}\boldsymbol{\Sigma}^{(b-1)}\ge\mathbf{0}_N,
\end{eqnarray}
\end{subequations}
where $\bar{\mathbf{H}}^{(b)}_v=\sum_{m=1}^MP^{(b)}_m\mathbf{d}\big(\theta^{(b)}_m\big)\mathbf{d}^H\big(\theta^{(b)}_m\big)+\boldsymbol{\Sigma}^{(b-1)}$, $\bar{\mathbf{q}}^{(b)}_m=\big[\bar{\mathbf{H}}^{(b)}_v\big]^{-1}\mathbf{d}\big(\theta_m^{(b)}\big)P^{(b)}_m$, and the conditional probability density functions of $f_m(t)$ and $\mathbf{z}(t)$ can be derived from \cite{bibitem23}.

At the additional M-step, the algorithm estimates $\mathbf{P}$ and $\boldsymbol{\sigma}$ by maximizing \eqref{eq37}, which results in the $M+N$ parallel subproblems:
\begin{subequations}
\begin{eqnarray}       
\min_{P_m\ge0}\ln(P_m)+\hat{P}^{(b)}_m/P_m,\forall m,
\end{eqnarray}
\begin{eqnarray}       
\min_{\sigma_n>0}\ln(\sigma_n)+\big[\hat{\mathbf{R}}^{(b)}_z\big]_{n,n}/\sigma_n,\forall n.
\end{eqnarray}
\end{subequations}
Thus, the estimates of $\mathbf{P}$ and $\boldsymbol{\sigma}$ are updated by
\begin{subequations}
\begin{eqnarray}       
P^{(b)}_m=\hat{P}^{(b)}_m,\forall m,
\end{eqnarray}
\begin{eqnarray}\label{eq40b}       
\sigma^{(b)}_n=\big[\hat{\mathbf{R}}^{(b)}_z\big]_{n,n}\ge0,\forall n,
\end{eqnarray}
\end{subequations}
which indicate that the estimate of $\mathbf{P}$ is updated again at this iteration and $\sigma^{(b)}_n=0$ is possible although its probability is very low. For example, if $M=1$, $T=1$, and $P^{(1)}_1=0$ in \eqref{eq34b}, we will have
$\bar{\mathbf{H}}^{(1)}_v=\boldsymbol{\Sigma}^{(0)}>\mathbf{0}_N$
and $\hat{\mathbf{R}}^{(1)}_z=\mathbf{v}(1)\mathbf{v}^H(1)$ in \eqref{eq38b}. Furthermore, if $\big[\mathbf{v}(1)\big]_n=0$, we will obtain $\sigma^{(1)}_n=0$ by \eqref{eq40b}. To avoid $\sigma^{(b)}_n=0$, we require that if $\sigma^{(b)}_n=0$ in \eqref{eq40b}, $\sigma^{(b)}_n$ will be updated by
\begin{eqnarray}       
\sigma^{(b)}_n=\zeta\sigma^{(b-1)}_n+(1-\zeta)\big[\hat{\mathbf{R}}^{(b)}_z\big]_{n,n},\nonumber
\end{eqnarray}
where $\zeta\in(0,1]$ and $\sigma^{(b)}_n\ge\zeta\sigma^{(b-1)}_n>0$ if $\sigma^{(b-1)}_n>0$, i.e., $\boldsymbol{\sigma}^{(b)}>\mathbf{0}$ if $\boldsymbol{\sigma}^{(b-1)}>\mathbf{0}$.

After the additional M-step, this iteration finishes and following \eqref{eq35}, we can verify the monotonicity of the first SAGE algorithm by
\begin{eqnarray}
\mathcal{L}\big(\boldsymbol{\Phi}^{(b)},\boldsymbol{\sigma}^{(b)}\big)
\ge\mathcal{L}\big(\boldsymbol{\Phi}^{(b)},\boldsymbol{\sigma}^{(b-1)}\big)
\ge\mathcal{L}\big(\boldsymbol{\Phi}^{(b-1)},\boldsymbol{\sigma}^{(b-1)}\big).
\end{eqnarray}

\subsection{Second SAGE Algorithm}

The first SAGE algorithm updates the DOA estimates simultaneously in \eqref{eq34a}, which leads to slow convergence. In order to speed up the convergence, we present the second SAGE algorithm, which updates the DOA estimates sequentially from $\theta_1$ to $\theta_M$ like the SAGE algorithm in Subsection \ref{sub:32}.

When the second SAGE algorithm updates the estimate of $\theta_i$ at the $i$th cycle of the $b$th iteration, all of the noise is also allocated to the $i$th source signal component. According to \eqref{eq15}, $\mathbf{g}_i(t)\sim\mathcal{CN}(\textbf{0},\mathbf{H}_i)$ with $\mathbf{H}_i=P_i\mathbf{d}(\theta_i)\mathbf{d}^H(\theta_i)+\boldsymbol{\Sigma}$ but for $m\ne i$, the statistical distribution of $\mathbf{g}_m(t)$ depends only on $f_m(t)$. The corresponding complete-data LLF is written as
\begin{eqnarray}\label{eq42}       
\mathcal{U}(\theta_i,\mathbf{P},\boldsymbol{\sigma})=\sum_{t=1}^T\big[\sum_{m\ne i}\ln\mathnormal{p}(f_m(t);P_m)+\ln\mathnormal{p}(\mathbf{g}_i(t);\theta_i,P_i,\boldsymbol{\sigma})\big]\nonumber \\
=-T(M-1)\ln(\pi)-T\sum_{m\ne i}\big[\ln(P_m)+\hat{P}_m/P_m\big]
-TN\ln(\pi)-T\big[\ln\big(\mathrm{Det}(\mathbf{H}_i)\big)+\mathrm{Tr}\big(\mathbf{H}_i^{-1}\hat{\mathbf{R}}_i\big)\big],
\end{eqnarray}
which is unsuitable for updating the estimate of $\boldsymbol{\sigma}$ due to the complexity of $\ln\big(\mathrm{Det}(\mathbf{H}_i)\big)+\mathrm{Tr}\big(\mathbf{H}_i^{-1}\hat{\mathbf{R}}_i\big)$. To proceed, we hold $\boldsymbol{\sigma}=\boldsymbol{\sigma}^{(b-1)}$ fixed and let the algorithm only update the estimates of $\theta_i$ and $\mathbf{P}$ by the following E- and M-steps.

\subsubsection{E-step}

The algorithm calculates the conditional expectation of $\mathcal{U}(\theta_i,\mathbf{P},\boldsymbol{\sigma})$, i.e.,
\begin{eqnarray}\label{eq43}       
\mathbb{E}\left\{\mathcal{U}(\theta_i,\mathbf{P},\boldsymbol{\sigma})\big|\mathbf{V};\boldsymbol{\Phi}^{(b,i-1)},\boldsymbol{\sigma}^{(b-1)}\right\}
=-T(M-1)\ln(\pi)-TN\ln(\pi)\nonumber\\
-T\sum_{m\ne i}\big[\ln(P_m)+\hat{P}^{(b,i)}_m/P_m\big]
-T\big[\ln\big(\mathrm{Det}(\mathbf{H}_i)\big)+\mathrm{Tr}\big(\mathbf{H}_i^{-1}\hat{\mathbf{R}}^{(b)}_i\big)\big],
\end{eqnarray}
where 
\begin{eqnarray}       
\hat{P}^{(b,i)}_m=&\mathbb{E}\big\{\hat{P}_m\big|\mathbf{V};\boldsymbol{\Phi}^{(b,i-1)},\boldsymbol{\sigma}^{(b-1)}\big\}\nonumber\\
=&P^{(b,i-1)}_m\big[1-\mathbf{d}^H(\theta^{(b,i-1)}_m)\bar{\mathbf{q}}^{(b,i-1)}_m\big]
+[\bar{\mathbf{q}}^{(b,i-1)}_m]^H\hat{\mathbf{R}}_v\bar{\mathbf{q}}^{(b,i-1)}_m\ge0
\end{eqnarray}
with $\bar{\mathbf{H}}^{(b,i-1)}_v=\sum_{m=1}^MP_m^{(b,i-1)}\mathbf{d}(\theta^{(b,i-1)}_m)\mathbf{d}^H(\theta^{(b,i-1)}_i)+\boldsymbol{\Sigma}^{(b-1)}$ and
$\bar{\mathbf{q}}^{(b,i-1)}_m=\big[\bar{\mathbf{H}}^{(b,i-1)}_v\big]^{-1}\mathbf{d}\big(\theta^{(b,i-1)}_m\big)P^{(b,i-1)}_m$,
\begin{eqnarray}       
\hat{\mathbf{R}}^{(b)}_i=&\hat{\mathbf{R}}^{(b,i)}_i=\mathbb{E}\big\{\hat{\mathbf{R}}_i\big|\mathbf{V};\boldsymbol{\Phi}^{(b,i-1)},\boldsymbol{\sigma}^{(b-1)}\big\}\nonumber\\
=&\bar{\mathbf{H}}^{(b,i-1)}_i[\bar{\mathbf{H}}^{(b,i-1)}_v]^{-1}\hat{\mathbf{R}}_v[\bar{\mathbf{H}}^{(b,i-1)}_v]^{-1}\bar{\mathbf{H}}^{(b,i-1)}_i+
\bar{\mathbf{H}}^{(b,i-1)}_i-\bar{\mathbf{H}}^{(b,i-1)}_i[\bar{\mathbf{H}}^{(b,i-1)}_v]^{-1}\bar{\mathbf{H}}^{(b,i-1)}_i\ge\mathbf{0}_N
\end{eqnarray}
with $\bar{\mathbf{H}}^{(b,i-1)}_i=P_i^{(b,i-1)}\mathbf{d}(\theta^{(b,i-1)}_i)\mathbf{d}^H(\theta^{(b,i-1)}_i)+\boldsymbol{\Sigma}^{(b-1)}$.

\subsubsection{M-step}

The algorithm estimates $\theta_i$ and $\mathbf{P}$ by maximizing \eqref{eq43}, i.e.,
\begin{eqnarray}       
\min_{\theta_i\in(0,\pi),\mathbf{P}\ge0}\sum_{m\ne i}\big[\ln(P_m)+\hat{P}^{(b,i)}_m/P_m\big]
+\big[\ln\big(\mathrm{Det}(\mathbf{H}_i)\big)+\mathrm{Tr}\big(\mathbf{H}_i^{-1}\hat{\mathbf{R}}^{(b)}_i\big)\big],
\end{eqnarray}
which leads to that
\begin{eqnarray}       
P^{(b,i)}_m=\hat{P}^{(b,i)}_m,\forall m\ne i,
\end{eqnarray}
while the estimates of $\theta_i$ and $P_i$ are updated by
\begin{eqnarray}       
\min_{\theta_i\in(0,\pi),P_i\ge0}\ln\big(\mathrm{Det}(\mathbf{K}_i)\big)+\mathrm{Tr}\big(\mathbf{K}_i^{-1}\widetilde{\mathbf{R}}^{(b)}_i\big),
\end{eqnarray}
where $\mathbf{H}_i=\big[\boldsymbol{\Sigma}^{(b-1)}\big]^{1/2}\mathbf{K}_i\big[\boldsymbol{\Sigma}^{(b-1)}\big]^{1/2}$,
$\mathbf{K}_i=P_i\tilde{\mathbf{d}}(\theta_i)\tilde{\mathbf{d}}^H(\theta_i)+\mathbf{I}_N$,
$\tilde{\mathbf{d}}(\theta_i)=\big[\boldsymbol{\Sigma}^{(b-1)}\big]^{-1/2}\mathbf{d}(\theta_i)$,
$\widetilde{\mathbf{R}}^{(b)}_i=\big[\boldsymbol{\Sigma}^{(b-1)}\big]^{-1/2}\hat{\mathbf{R}}^{(b)}_i\big[\boldsymbol{\Sigma}^{(b-1)}\big]^{-1/2}$.
Following \eqref{eq34a}--\eqref{eq34b}, the estimates of $\theta_i$ and $P_i$ can be updated by
\begin{subequations}
\begin{eqnarray}       
\theta^{(b)}_i=\theta^{(b,i)}_i=\arg\max_{\theta_i\in(0,\pi)}\tilde{\mathbf{d}}^H(\theta_i)\widetilde{\mathbf{R}}^{(b)}_i\tilde{\mathbf{d}}(\theta_i),
\end{eqnarray}
\begin{eqnarray}       
P^{(b,i)}_i=\max\bigg\{\frac{1}{q^{(b)}}\Big(\frac{\tilde{\mathbf{d}}^H(\theta^{(b)}_i)\widetilde{\mathbf{R}}^{(b)}_i\tilde{\mathbf{d}}(\theta^{(b)}_i)}{q^{(b)}}-1\Big),0\bigg\}.
\end{eqnarray}
\end{subequations}

The other signal parameter estimate(s) in $\boldsymbol{\Phi}^{(b,i-1)}=(\boldsymbol{\theta}^{(b,i-1)},\mathbf{P}^{(b,i-1)})$ is(are) not updated at this cycle and the iterative value(s) is(are) denoted by
\begin{eqnarray}     
\theta^{(b,i)}_m=\theta^{(b,i-1)}_m,\forall m\ne i.
\end{eqnarray}
After this cycle, we have
\begin{eqnarray}  
\mathcal{L}\big(\boldsymbol{\Phi}^{(b,i)},\boldsymbol{\sigma}^{(b-1)}\big)\ge\mathcal{L}\big(\boldsymbol{\Phi}^{(b,i-1)},\boldsymbol{\sigma}^{(b-1)}\big).
\end{eqnarray}

The above E- and M-steps (or cycle) are repeated until the estimate of $\theta_M$ is updated and we further have
\begin{eqnarray}\label{eq52} 
\mathcal{L}\big(\boldsymbol{\Phi}^{(b)},\boldsymbol{\sigma}^{(b-1)}\big)=\mathcal{L}\big(\boldsymbol{\Phi}^{(b,M)},\boldsymbol{\sigma}^{(b-1)}\big)
\ge\cdots
\ge\mathcal{L}\big(\boldsymbol{\Phi}^{(b,0)},\boldsymbol{\sigma}^{(b-1)}\big)=\mathcal{L}\big(\boldsymbol{\Phi}^{(b-1)},\boldsymbol{\sigma}^{(b-1)}\big).
\end{eqnarray}

\subsubsection{Additional E- and M-steps}

The above procedure does not update the estimate of $\boldsymbol{\sigma}$, so we also use the additional E- and M-steps \ref{sec:413} of the first SAGE algorithm to obtain $\boldsymbol{\sigma}^{(b)}$ easily. After the additional E- and M-steps, the $b$th iteration finishes and following \eqref{eq52}, we can verify the monotonicity of the second SAGE algorithm by
\begin{eqnarray} 
\mathcal{L}\big(\boldsymbol{\Phi}^{(b)},\boldsymbol{\sigma}^{(b)}\big)
\ge\mathcal{L}\big(\boldsymbol{\Phi}^{(b)},\boldsymbol{\sigma}^{(b-1)}\big)
\ge\mathcal{L}\big(\boldsymbol{\Phi}^{(b-1)},\boldsymbol{\sigma}^{(b-1)}\big).
\end{eqnarray}

\section{Convergence Properties of the EM-Type Algorithms}

\subsection{Convergence Point}

It is easy to verify that the above EM-type algorithms satisfy standard regularity conditions and the two algorithms for the same signal model always converge to different stationary points or the same stationary point of $\mathcal{L}(\boldsymbol{\Phi},\boldsymbol{\sigma})$ \cite{bibitem6}, \cite{bibitem9}, \cite{bibitem24}.

It is well known that these algorithms, as ``hill climbing'' algorithms, require accurate initial points for obtaining their global maximum points. Moreover, note that processing the same samples, the two algorithms for the same signal model may converge to different stationary points of $\mathcal{L}(\boldsymbol{\Phi},\boldsymbol{\sigma})$ given the same initial point, which indicates that one of the two algorithm may be more efficient for avoiding the convergence to an unwanted stationary point of $\mathcal{L}(\boldsymbol{\Phi},\boldsymbol{\sigma})$ than the other. Hence, we compare convergence points of the EM-type algorithms given poor initial points in the next section.

\subsection{Complexity and Stability}

The computational complexities of the EM-type algorithms at the $b$th iteration mainly lie in the $M$ one-dimensional search problems:
\begin{eqnarray}\label{eq54}       
\theta^{(b)}_m=\arg\max_{\theta_m\in(0,\pi)}\tilde{\mathbf{d}}^H(\theta_m)\widetilde{\mathbf{R}}^{(b)}_m\tilde{\mathbf{d}}(\theta_m),\forall m.
\end{eqnarray}
However, we find that when the powers of sources are unequal, the DOA estimates of multiple sources, updated by \eqref{eq54}, tend to be consistent with the true DOA of the source with the largest power. As a consequence, the EM-type algorithms are unstable.

To address this issue, we reduce the difference between $\theta_m^{(b)}$ and $\theta_m^{(b-1)}$ and still use the method in our previous works \cite{bibitem12}, \cite{bibitem22}, i.e., choosing $\theta_m^{(b-1)}$ as the initial point of some gradient algorithm and then applying this gradient algorithm to search a local maximum point of $\tilde{\mathbf{d}}^H(\theta_m)\widetilde{\mathbf{R}}^{(b)}_m\tilde{\mathbf{d}}(\theta_m)$ as $\theta_m^{(b)}$, which leads to \textbf{Algorithm 1} in the next section. Using this method, we still have
\begin{eqnarray}\label{eq55} 
\tilde{\mathbf{d}}^H(\theta^{(b)}_m)\widetilde{\mathbf{R}}^{(b)}_m\tilde{\mathbf{d}}(\theta^{(b)}_m)\ge
\tilde{\mathbf{d}}^H(\theta^{(b-1)}_m)\widetilde{\mathbf{R}}^{(b)}_m\tilde{\mathbf{d}}(\theta^{(b-1)}_m),\forall m,
\end{eqnarray}
which guarantees the monotonicity of the EM-type algorithms.

\section{Simulation Results}

Simulation results are given to illustrate the convergence of the EM-type algorithms. We adopt $M=2$ and $\Vert\boldsymbol{\theta}^{(b)}-\boldsymbol{\theta}^{(b-1)}\Vert\le0.001^{\circ}$ as the stopping criterion of the EM-type algorithms. $\mathbf{F}$ in the deterministic signal model is also generated by the independent random numbers $f_m(t)\sim\mathcal{CN}(0,P_m)$. \textbf{Algorithm 1} is used to search the $\theta^{(b)}_m$'s in \eqref{eq55}. Moreover, $N=10$ and $\boldsymbol{\sigma}=[1.1~2.3~3~4.2~1.3~0.5~5~2.2~6.7~10]^T$.

\begin{algorithm}
\caption{Gradient Ascent Based Angle Estimation} \label{alg:A}
\begin{algorithmic}[1]
\STATE {$h(\theta_m)=\tilde{\mathbf{d}}^H(\theta_m)\widetilde{\mathbf{R}}^{(b)}_m\tilde{\mathbf{d}}(\theta_m)$, initialize $\theta_m=\theta_m^{(b-1)}\in(0,\pi)$.}
        \WHILE{$\vert h'(\theta_m)\vert>0.001$}
           \STATE {$t=0.1\times\left\{
\begin{array}{ll}
(\pi-\theta_m)/h'(\theta_m),&h'(\theta_m)>0,\\
-\theta_m/h'(\theta_m),&h'(\theta_m)<0.
\end{array}
\right.$}
           \WHILE{$h\big(\theta_m+th'(\theta_m)\big)<h(\theta_m)+0.3t\vert h'(\theta_m)\vert^2$}
           \STATE {$t=0.5t$.}
           \ENDWHILE
           \STATE {$\theta_m=\theta_m+th'(\theta_m)$.}
       \ENDWHILE
\STATE {$\theta_m^{(b)}=\theta_m$.}
\end{algorithmic}
\end{algorithm}

\subsection{Deterministic Signal Model}

Fig. 1 plots the $\mathcal{L}\big(\boldsymbol{\Phi}^{(b)},\boldsymbol{\sigma}^{(b)}\big)$'s, $\theta^{(b)}_1$'s, and $\theta^{(d)}_2$'s obtained by the GEM and SAGE algorithms under one realization without loss of generality. Both algorithms share the same initial point and process the same samples. It is straightforward to see that given a good initial point, both algorithms obtain consistent DOA estimates and the SAGE algorithm has faster convergence than the GEM algorithm.

\begin{figure}[t] \centering
\includegraphics[scale=0.8]{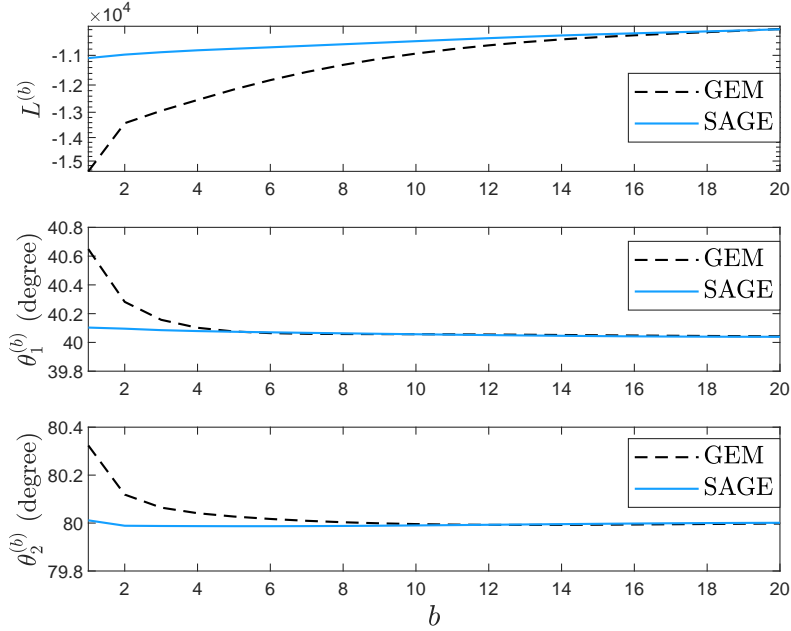}\label{f1}
\vspace{0cm}\caption{$\mathcal{L}\big(\boldsymbol{\Phi}^{(b)},\boldsymbol{\sigma}^{(b)}\big)$, $\theta^{(b)}_1$, and $\theta^{(b)}_2$ comparison of the GEM and SAGE algorithms under one realization with $\beta=0.5$, $\gamma=0.9$, $T=500$, $\theta_1=40^{\circ}$, $\theta_2=80^{\circ}$, $P_1=6$, $P_2=8$, $\theta^{(0)}_1=45^{\circ}$, $\theta^{(0)}_2=85^{\circ}$, $\mathbf{F}^{(0)}=[\mathbf{1}~\mathbf{1}]^T$, $\boldsymbol{\Omega}^{(0)}=(\mathbf{1},\mathbf{1})/2$, and $\boldsymbol{\sigma}^{(0)}=\mathbf{1}$.}\vspace{0cm}
\end{figure}

Fig. 2 shows a scatter plot of the DOA estimates obtained by both algorithms under 100 independent realizations. Given the same initial point, the same samples of each realization are processed by both algorithms. In Fig. 2, the total numbers of wanted points from the GEM and SAGE algorithms are 8 and 100, respectively. Hence, we conclude that given a poor initial point, the SAGE algorithm can avoid the convergence to an unwanted stationary point of $\mathcal{L}(\boldsymbol{\Phi},\boldsymbol{\sigma})$ more efficiently than the GEM algorithm.

\begin{figure}[t] \centering
\includegraphics[scale=0.8]{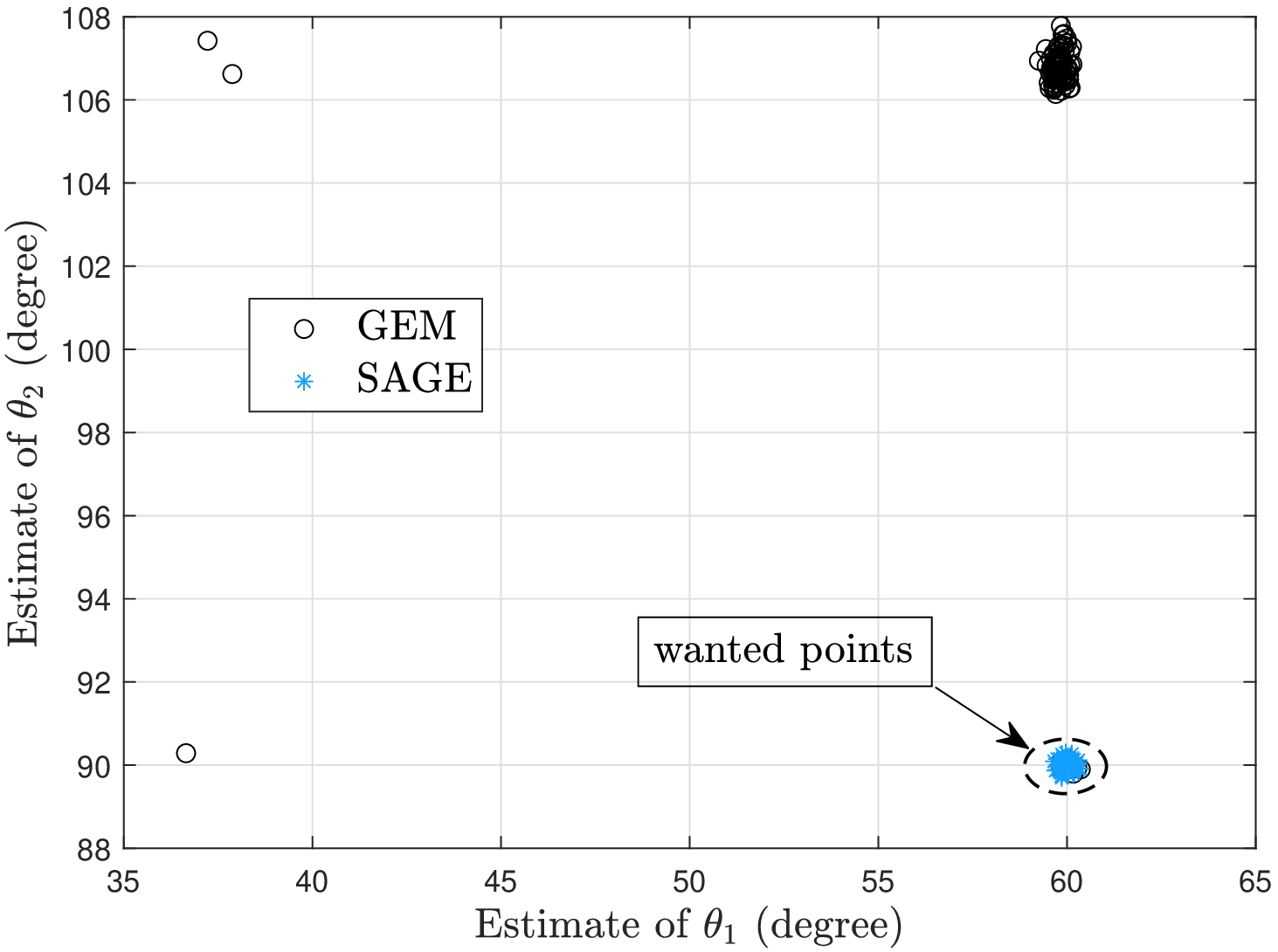}\label{f2}
\vspace{0cm}\caption{Scatter plot of the DOA estimates obtained by the GEM and SAGE algorithms under 100 independent realizations with $\beta=0.95$, $\gamma=0.99$, $T=150$, $\theta_1=60^{\circ}$, $\theta_2=90^{\circ}$, $P_1=P_2=3$, $\theta^{(0)}_1=40^{\circ}$, $\theta^{(0)}_2=110^{\circ}$, $\textbf{F}^{(0)}=[\textbf{1}~\textbf{1}]^T$, $\boldsymbol{\Omega}^{(0)}=(\mathbf{1},\mathbf{1})/2$, and $\boldsymbol{\sigma}^{(0)}=\mathbf{1}$.}\vspace{0cm}
\end{figure}

According to Figs. 1 and 2, we can conclude that \emph{for the deterministic signal model, the SAGE algorithm outperforms the GEM algorithm}. Thus, we only use the SAGE algorithm in Figs. 3 and 4.

Figs. 3 and 4 compare the root mean square error (RMSE) performances of DOA estimation obtained by the SAGE algorithm for the nonuniform and uniform noise models. The SAGE algorithm for unknown uniform noise is presented in \cite{bibitem22}. The SAGE algorithm for each noise model is performed based on a good initial point for avoiding the convergence to an unwanted stationary point efficiently. Each RMSE in Figs. 3 and 4 is computed from 1000 independent realizations with the same $\mathbf{F}$ and the corresponding Cramer-Rao lower bound (CRLB) is also provided \cite{bibitem19}. From Figs. 3 and 4, we can observe that as expected, the SAGE algorithm for nonuniform noise yields smaller RMSEs than that for uniform noise. Moreover, we note that the SAGE algorithm for nonuniform noise cannot achieve the CRLB of $\boldsymbol{\theta}$ by increasing $T$ or $P$, which is consistent with a main conclusion in \cite{bibitem4}, i.e., \emph{the deterministic ML estimator of $\boldsymbol{\theta}$ is not statistically efficient if the number of sensors $N$ is small}.

\begin{figure}[t] \centering
\includegraphics[scale=0.8]{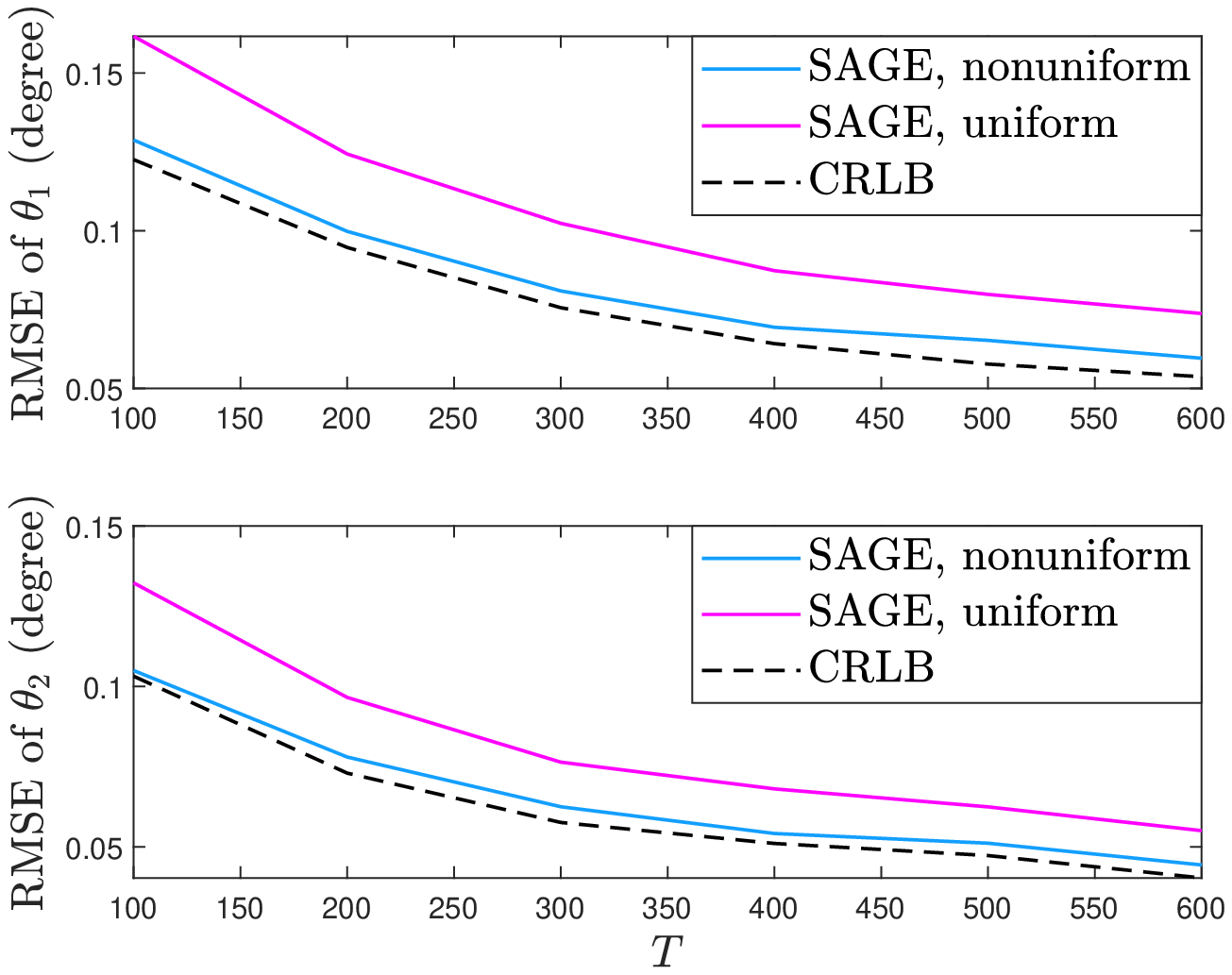}
\vspace{0cm}\caption{RMSEs of DOA estimation obtained by the SAGE algorithm for the nonuniform and uniform noise models with $\gamma=0.99$, $\theta_1=45^{\circ}$, $\theta_2=65^{\circ}$, $P_1=P_2=7$, $\theta^{(0)}_1=40^{\circ}$, $\theta^{(0)}_2=70^{\circ}$, $\textbf{F}^{(0)}=[\textbf{1}~\textbf{1}]^T$, $\boldsymbol{\sigma}^{(0)}=\mathbf{1}$, and $\sigma^{(0)}=1$.}\vspace{0cm}
\end{figure}

\begin{figure}[t] \centering
\includegraphics[scale=0.8]{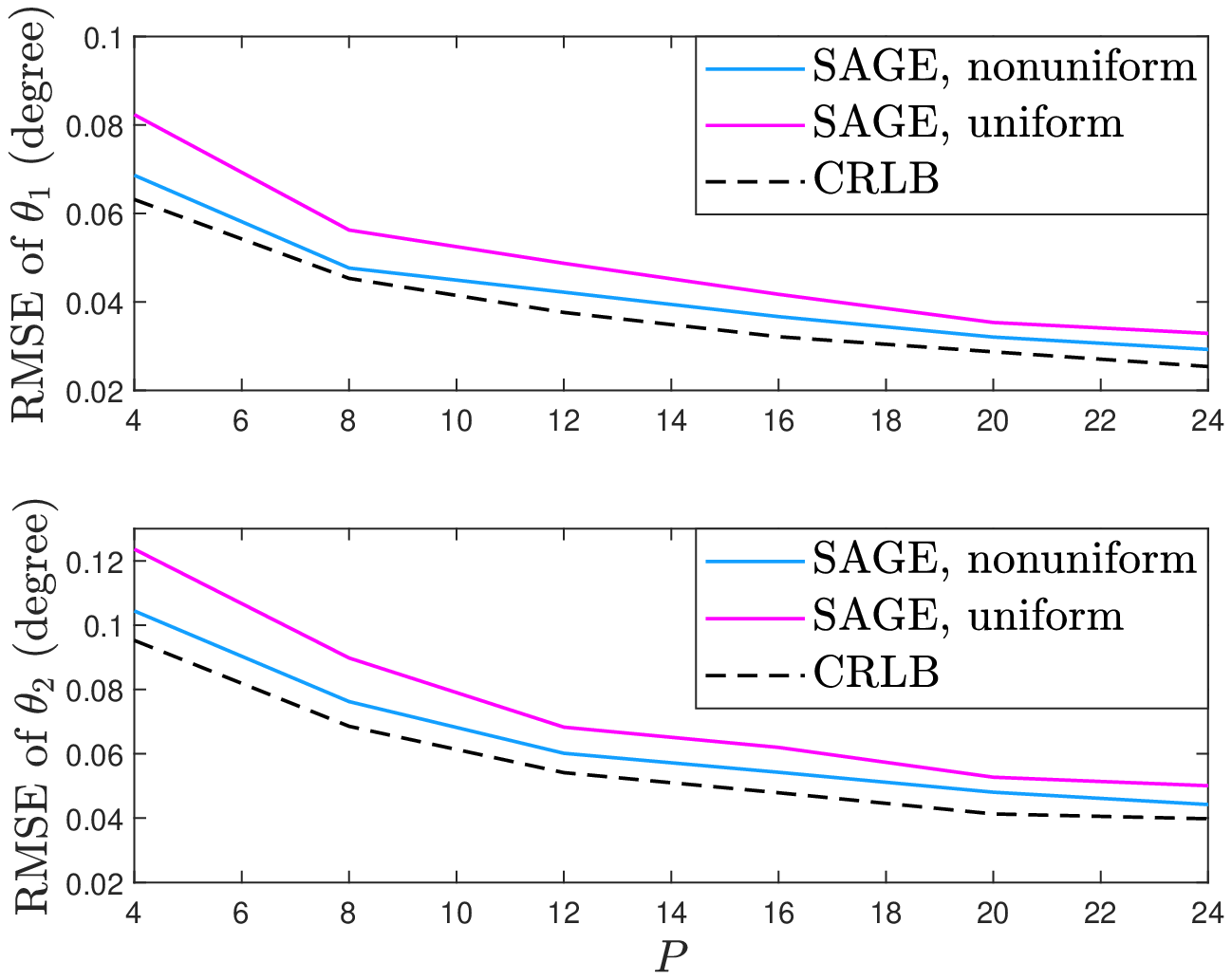}
\vspace{0cm}\caption{RMSEs of DOA estimation obtained by the SAGE algorithm for the nonuniform and uniform noise models with $\gamma=0.99$, $\theta_1=80^{\circ}$, $\theta_2=140^{\circ}$, $T=400$, $P_1=P_2=P$, $\theta^{(0)}_1=75^{\circ}$, $\theta^{(0)}_2=135^{\circ}$, $\textbf{F}^{(0)}=[\textbf{1}~\textbf{1}]^T$, $\boldsymbol{\sigma}^{(0)}=\mathbf{1}$, and $\sigma^{(0)}=1$.}\vspace{0cm}
\end{figure}

\subsection{Stochastic Signal Model}

Fig. 5 plots the $\mathcal{L}\big(\boldsymbol{\Phi}^{(b)},\boldsymbol{\sigma}^{(b)}\big)$'s, $\theta^{(b)}_1$'s, and $\theta^{(b)}_2$'s obtained by the first and second SAGE algorithms under one realization. Both algorithms share the same initial point and process the same samples. It is straightforward to observe that given a good initial point, both algorithms obtain consistent DOA estimates and the second SAGE algorithm has faster convergence than the first SAGE algorithm.

\begin{figure}[t] \centering
\includegraphics[scale=0.8]{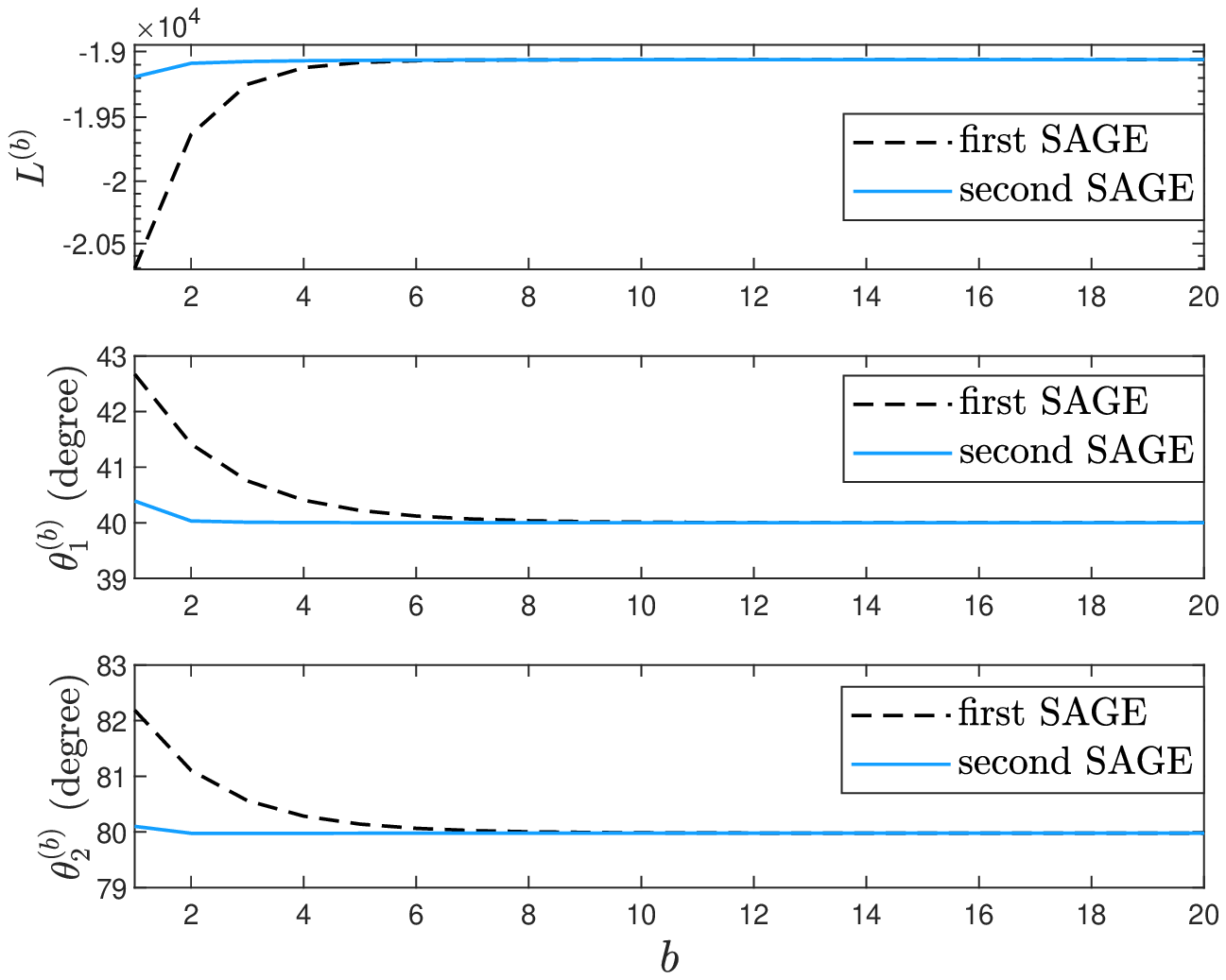}
\vspace{0cm}\caption{$\mathcal{L}\big(\boldsymbol{\Phi}^{(b)},\boldsymbol{\sigma}^{(b)}\big)$, $\theta^{(b)}_1$, and $\theta^{(b)}_2$ comparison of the first and second SAGE algorithms under one realization with $\boldsymbol{\alpha}=[0.5~0.5]^T$, $T=500$, $\theta_1=40^{\circ}$, $\theta_2=80^{\circ}$, $P_1=6$, $P_2=8$, $\theta^{(0)}_1=45^{\circ}$, $\theta^{(0)}_2=85^{\circ}$, $\mathbf{P}^{(0)}=\mathbf{1}$, and $\boldsymbol{\sigma}^{(0)}=\mathbf{1}$.}\vspace{0cm}
\end{figure}

Fig. 6 shows a scatter plot of the DOA estimates obtained by both algorithms under 100 independent realizations. Given the same initial point, the same samples of each realization are processed by both algorithms. In Fig. 6, the total numbers of wanted points from the first and second SAGE algorithms are 90 and 100, respectively. Hence, we conclude that given a poor initial point, the second SAGE algorithm can avoid the convergence to an unwanted stationary point of $\mathcal{L}(\boldsymbol{\Phi},\boldsymbol{\sigma})$ more efficiently than the first SAGE algorithm.

\begin{figure}[t] \centering
\includegraphics[scale=0.8]{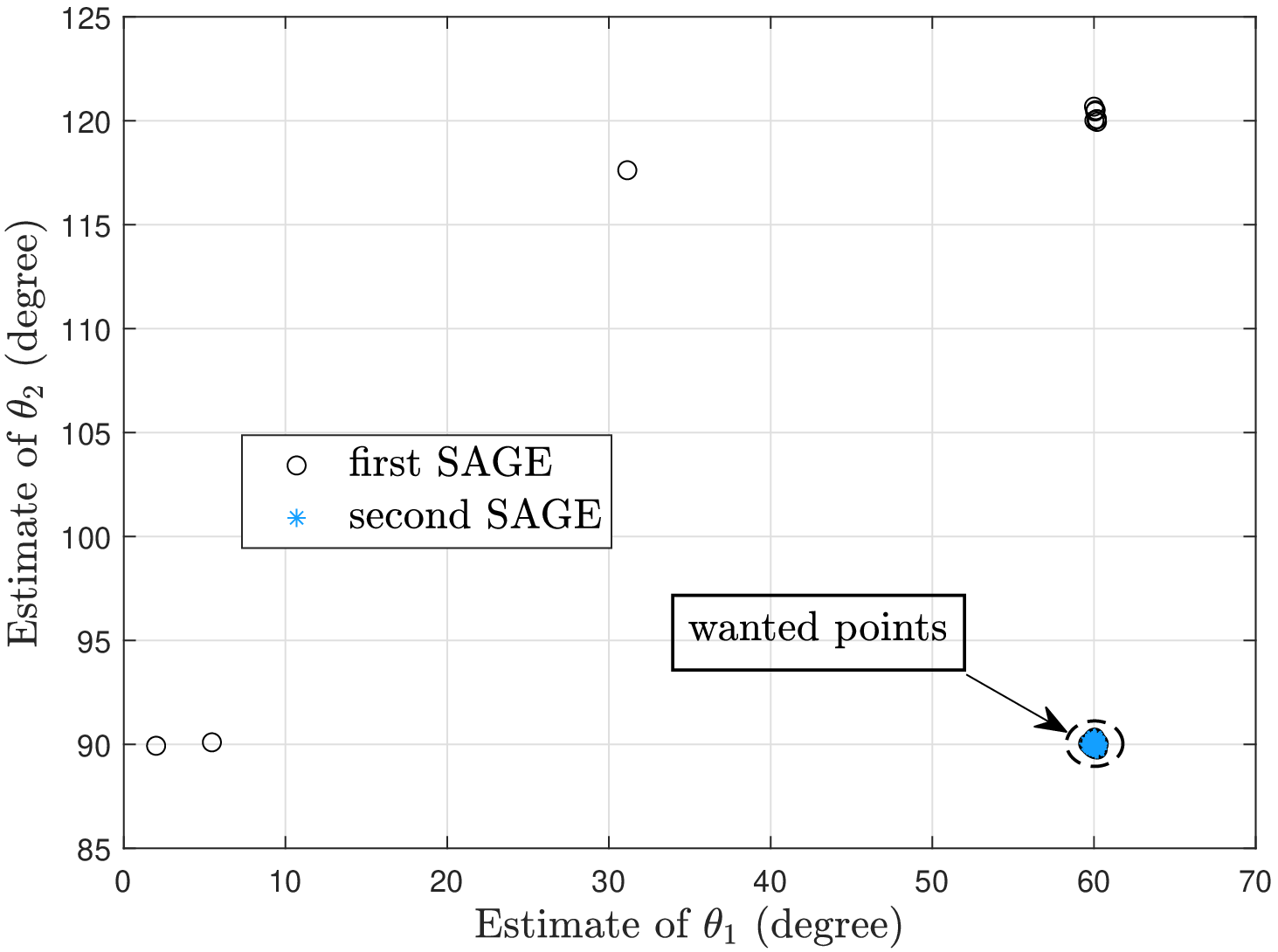}
\vspace{0cm}\caption{Scatter plot of the DOA estimates obtained by the first and second SAGE algorithms under 100 independent realizations with $\boldsymbol{\alpha}=[0.5~0.5]^T$, $T=150$, $\theta_1=60^{\circ}$, $\theta_2=90^{\circ}$, $P_1=P_2=3$, $\theta^{(0)}_1=40^{\circ}$, $\theta^{(0)}_2=110^{\circ}$, $\textbf{P}^{(0)}=\textbf{1}$, and $\boldsymbol{\sigma}^{(0)}=\mathbf{1}$.}\vspace{0cm}
\end{figure}

According to Figs. 5 and 6, we can conclude that \emph{for the stochastic signal model, the second SAGE algorithm outperforms the first SAGE algorithm}. Thus, we only use the second SAGE algorithm in Figs. 7--9.

Figs. 7--9 compare the RMSE performances of DOA estimation obtained by the SAGE algorithm for the nonuniform and uniform noise models. The SAGE algorithm for unknown uniform noise is presented in \cite{bibitem22}. The SAGE algorithm for each noise model is performed based on a good initial point for avoiding the convergence to an unwanted stationary point efficiently. Each RMSE in Figs. 7--9 is computed from 1000 independent realizations and the corresponding CRLB is provided \cite{bibitem19}. From Figs. 7--9, we can observe that as expected, the SAGE algorithm for nonuniform noise yields smaller RMSEs than that for uniform noise. Moreover, we note that the SAGE algorithm for nonuniform noise can achieve the CRLB of $\boldsymbol{\theta}$ by increasing $T$, which is consistent with a main conclusion in \cite{bibitem5}, i.e., \emph{the stochastic ML estimator of $\boldsymbol{\theta}$ asymptotically achieves the CRLB of $\boldsymbol{\theta}$ if the number of samples $T$ is large}.

\begin{figure}[t] \centering
\includegraphics[scale=0.8]{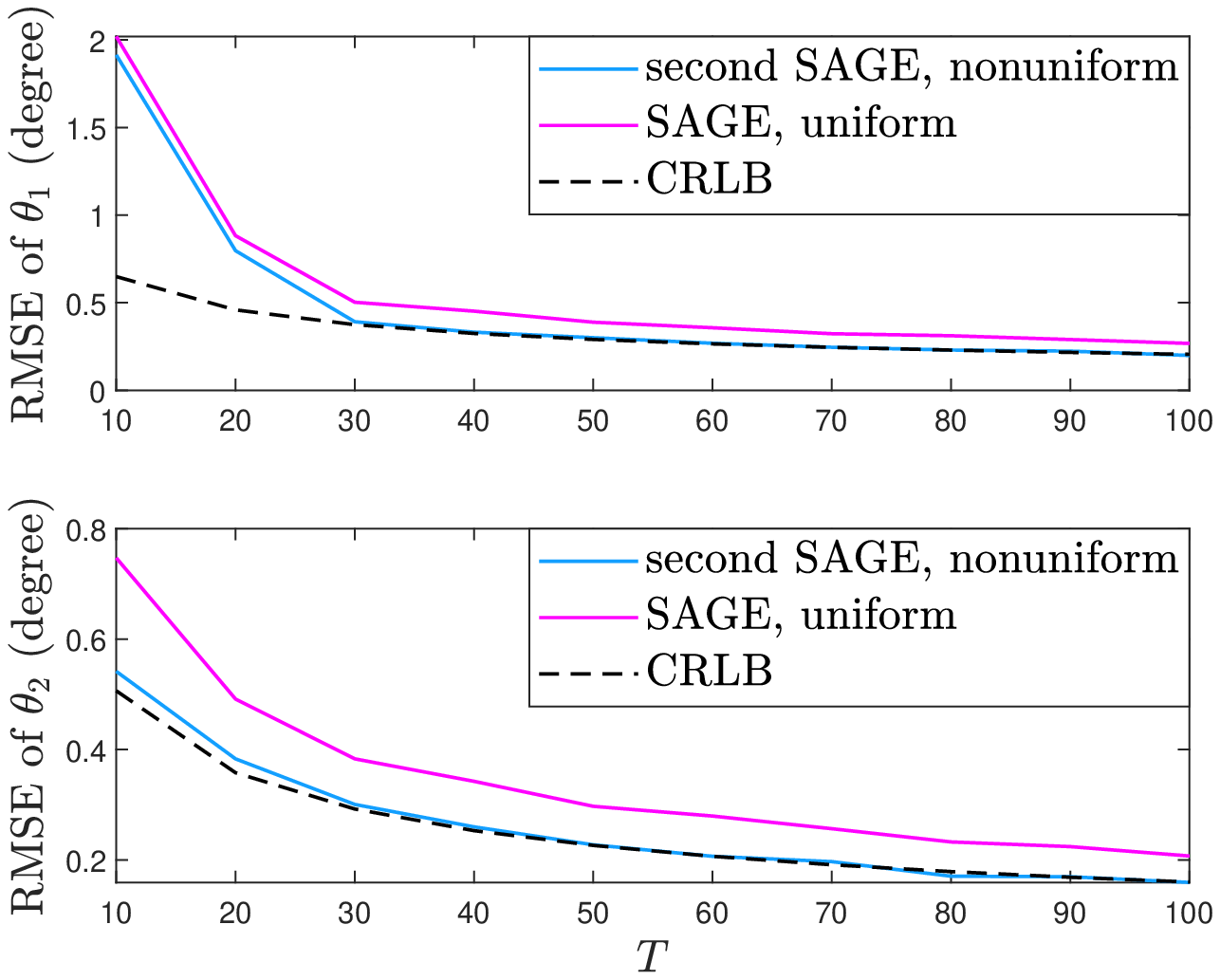}
\vspace{0cm}\caption{RMSEs of DOA estimation obtained by the SAGE algorithm for the nonuniform and uniform noise models with $\theta_1=45^{\circ}$, $\theta_2=65^{\circ}$, $P_1=P_2=3$, $\theta^{(0)}_1=40^{\circ}$, $\theta^{(0)}_2=70^{\circ}$, $\textbf{P}^{(0)}=\textbf{1}$, $\boldsymbol{\sigma}^{(0)}=\mathbf{1}$, and $\sigma^{(0)}=1$.}\vspace{0cm}
\end{figure}

\begin{figure}[t] \centering
\includegraphics[scale=0.8]{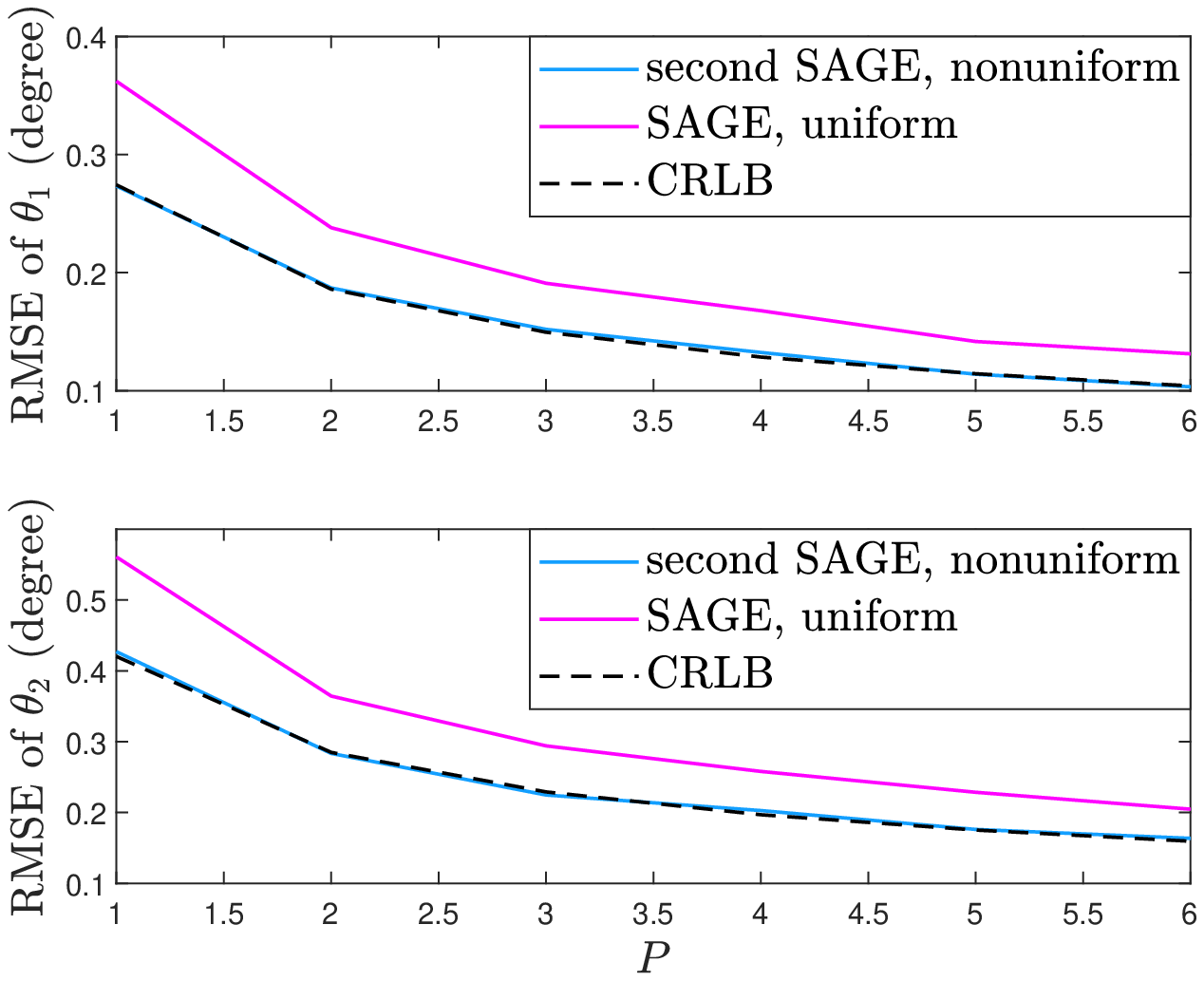}
\vspace{0cm}\caption{RMSEs of DOA estimation obtained by the SAGE algorithm for the nonuniform and uniform noise models with $\theta_1=80^{\circ}$, $\theta_2=140^{\circ}$, $T=100$, $P_1=P_2=P$, $\theta^{(0)}_1=75^{\circ}$, $\theta^{(0)}_2=135^{\circ}$, $\textbf{P}^{(0)}=\textbf{1}$, $\boldsymbol{\sigma}^{(0)}=\mathbf{1}$, and $\sigma^{(0)}=1$.}\vspace{0cm}
\end{figure}

\begin{figure}[t] \centering
\includegraphics[scale=0.8]{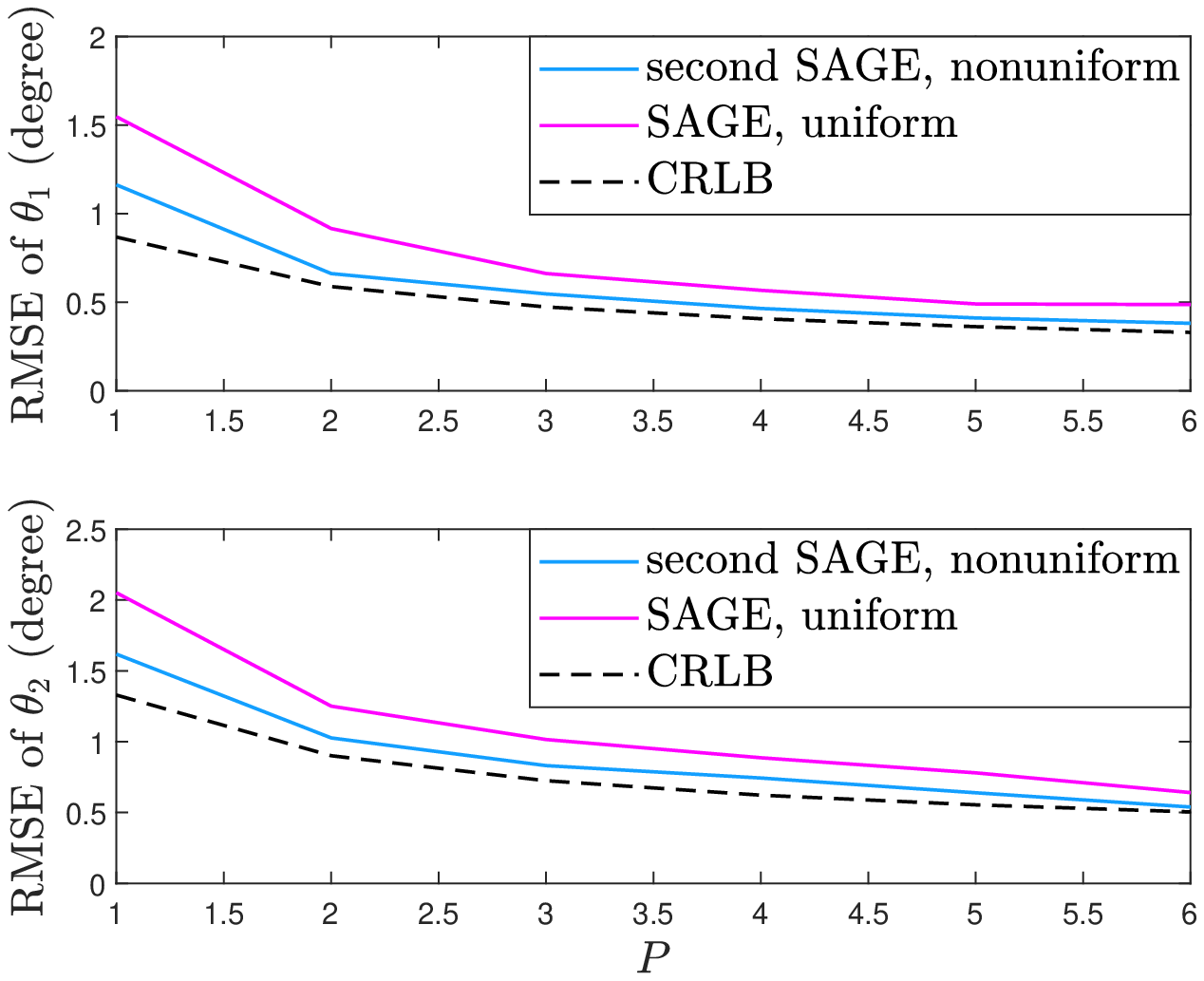}
\vspace{0cm}\caption{RMSEs of DOA estimation obtained by the SAGE algorithm for the nonuniform and uniform noise models with $\theta_1=80^{\circ}$, $\theta_2=140^{\circ}$, $T=10$, $P_1=P_2=P$, $\theta^{(0)}_1=75^{\circ}$, $\theta^{(0)}_2=135^{\circ}$, $\textbf{P}^{(0)}=\textbf{1}$, $\boldsymbol{\sigma}^{(0)}=\mathbf{1}$, and $\sigma^{(0)}=1$.}\vspace{0cm}
\end{figure}

\subsection{Deterministic and Stochastic Signal Models}

The SAGE algorithm for the deterministic signal model can process samples from the stochastic signal model, so we compare it with the second SAGE algorithm for the stochastic signal model in this subsection. Note that the two algorithms estimate the same DOA parameter $\boldsymbol{\theta}$, the stopping criterion $\Vert\boldsymbol{\theta}^{(b)}-\boldsymbol{\theta}^{(b-1)}\Vert\le0.001^{\circ}$ is suitable.

Fig. 10 shows a scatter plot of the DOA estimates obtained by both algorithms under 50 independent realizations. The same samples of each realization are processed by both algorithms and each algorithm is performed based on a good initial point for avoiding the convergence to an unwanted stationary point efficiently. From Fig. 10, we can observe that both algorithms tend to obtain inconsistent DOA estimates. Hence, we compare the RMSE performances of both algorithms in Figs. 11 and 12.

\begin{figure}[t] \centering
\includegraphics[scale=0.8]{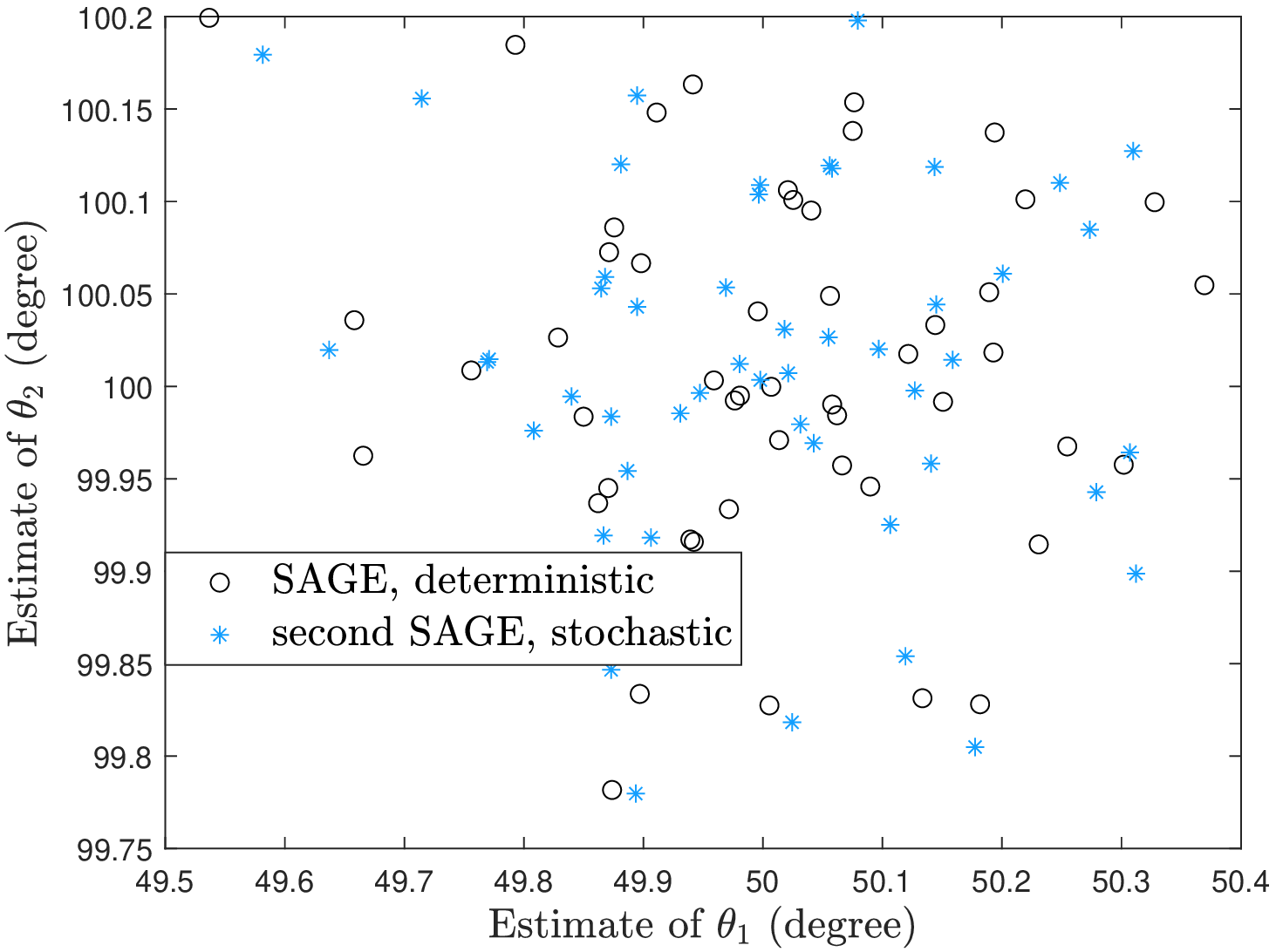}
\vspace{0cm}\caption{Scatter plot of the DOA estimates obtained by the SAGE algorithm for the deterministic and stochastic signal models under 50 independent realizations with $\gamma=0.99$, $T=100$, $\theta_1=50^{\circ}$, $\theta_2=100^{\circ}$, $P_1=4$, $P_2=6$, $\theta^{(0)}_1=45^{\circ}$, $\theta^{(0)}_2=105^{\circ}$, $\textbf{F}^{(0)}=[\textbf{1}~\mathbf{1}]^T$, $\textbf{P}^{(0)}=\textbf{1}$, and $\boldsymbol{\sigma}^{(0)}=\mathbf{1}$.}\vspace{0cm}
\end{figure}

Figs. 11 and 12 compare the RMSE performances of DOA estimation obtained by the SAGE algorithm for the deterministic and stochastic signal models. The SAGE algorithm for each signal model is performed based on a good initial point for avoiding the convergence to an unwanted stationary point efficiently. Each RMSE in Figs. 11 and 12 is computed from 1000 independent realizations. From Figs. 11 and 12, we can observe that the SAGE algorithm for the stochastic signal model yields smaller RMSEs than that for the deterministic signal model, which is consistent with a main conclusion in \cite{bibitem5}, i.e., \emph{the stochastic ML estimator of $\boldsymbol{\theta}$ is statistically more efficient than the deterministic ML estimator of $\boldsymbol{\theta}$}.

\begin{figure}[t] \centering
\includegraphics[scale=0.8]{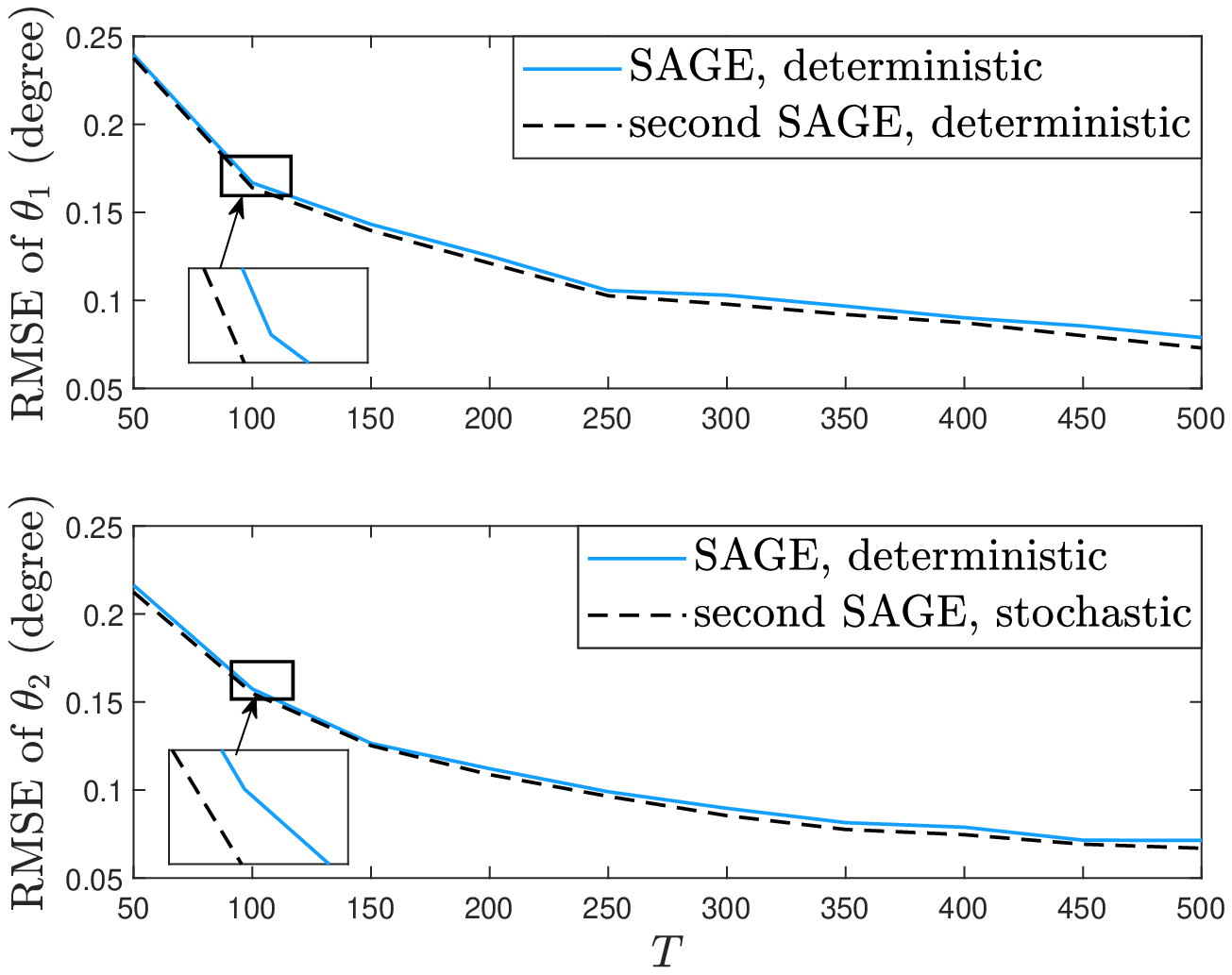}
\vspace{0cm}\caption{RMSEs of DOA estimation obtained by the SAGE algorithm for the deterministic and stochastic signal models with $\gamma=0.99$, $\theta_1=40^{\circ}$, $\theta_2=70^{\circ}$, $P_1=5$, $P_2=3$, $\theta^{(0)}_1=45^{\circ}$, $\theta^{(0)}_2=65^{\circ}$, $\textbf{F}^{(0)}=[\textbf{1}~\mathbf{1}]^T$, $\textbf{P}^{(0)}=\textbf{1}$, and $\boldsymbol{\sigma}^{(0)}=\mathbf{1}$.}\vspace{0cm}
\end{figure}

\begin{figure}[t] \centering
\includegraphics[scale=0.8]{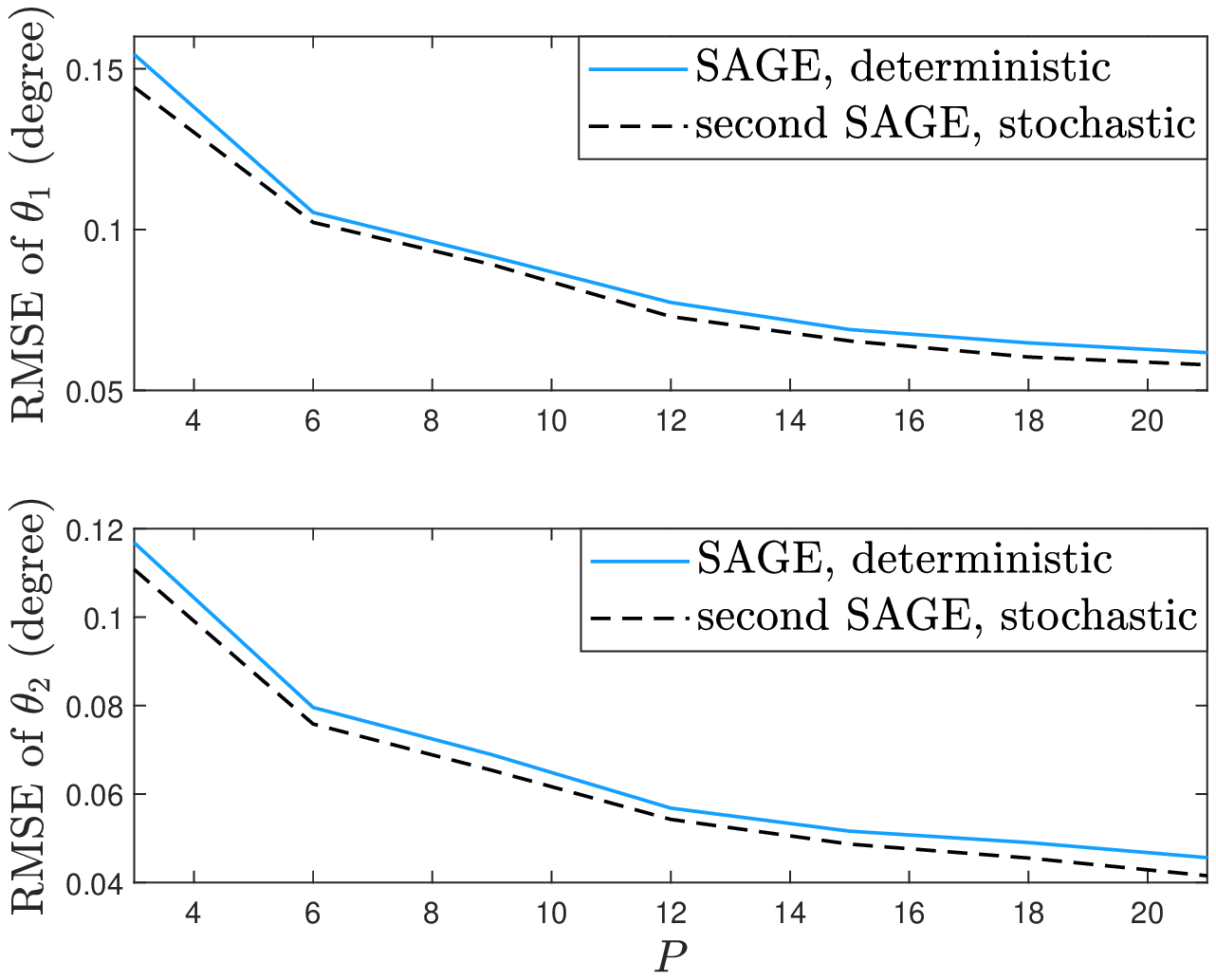}
\vspace{0cm}\caption{RMSEs of DOA estimation obtained by the SAGE algorithm for the deterministic and stochastic signal models with $\gamma=0.99$, $T=200$, $\theta_1=135^{\circ}$, $\theta_2=75^{\circ}$, $P_1=P_2=P$, $\theta^{(0)}_1=130^{\circ}$, $\theta^{(0)}_2=70^{\circ}$, $\textbf{F}^{(0)}=[\textbf{1}~\mathbf{1}]^T$, $\textbf{P}^{(0)}=\textbf{1}$, and $\boldsymbol{\sigma}^{(0)}=\mathbf{1}$.}\vspace{0cm}
\end{figure}

\section{Conclusion}

We have presented several EM-type algorithms for efficiently computing both the deterministic and stochastic ML estimators in unknown nonuniform noise. Specifically, we design a GEM algorithm and an SAGE algorithm for computing the deterministic ML estimator. Simulation results show that the SAGE algorithm converges faster and is more efficient for avoiding the convergence to an unwanted stationary point of the LLF. Moreover, we design two SAGE algorithms for computing the stochastic ML estimator, in which the first updates the DOA estimates simultaneously while the second updates the DOA estimates sequentially. Simulation results show that the second SAGE algorithm converges faster and is more efficient for avoiding the convergence to an unwanted stationary point of the LLF.


\end{document}